%% file: arxiv-orbit-main.tex
\documentclass[journal,unknownkeysallowed]{vgtc}

\onlineid{1844}

\vgtccategory{Research}

\vgtcpapertype{application/design study}

\usepackage{tabu}                      
\usepackage{booktabs}                  
\usepackage{flushend}
\usepackage{mathptmx}
\usepackage[dvipsnames,svgnames]{xcolor}
\usepackage{url}

\newcommand{\edit}[1]{{\textcolor{black}{#1}}}

\newcommand{\etal}{et~al.}
\newcommand{\ie}{{i.e.}}

\newcommand{\findorb}{\texttt{find\_orb}}
\newcommand{\adamcore}{\texttt{adam\_core}}
\newcommand{\mpcq}{\texttt{mpcq}}
\newcommand{\neoviz}{\emph{NEOviz}}

\newcommand{\para}[1]{\vspace{0pt}\noindent{\textbf{#1.}}\hspace{0.75em}}
\newcommand{\designation}[3]{#1 #2$_{\textrm{\small #3}}$}

\newcommand{\dateth}{$^{\textrm{\small th}}$}

\title{NEOviz: Uncertainty-Driven Visual Analysis of Asteroid Trajectories}

\author{
  \authororcid{Fangfei Lan}{0000-0002-8237-5919},
  \authororcid{Malin Ejdbo}{0009-0005-6870-8552},
  \authororcid{Joachim Moeyens}{0000-0001-5820-3925},
  \authororcid{Bei Wang}{0000-0002-9240-0700},
  \authororcid{Anders Ynnerman}{0000-0002-9466-9826},
  \authororcid{Alexander Bock}{0000-0002-2849-6146}
}

\authorfooter{
  \item Fangfei Lan and Malin Ejdbo contributed equally.
  \item
    Fangfei Lan and Bei Wang are with the University of Utah.\\
    E-mail: \{fangfei.lan, beiwang\}@sci.utah.edu
  \item
    Malin Ejdbo, Anders Ynnerman, and Alexander Bock are with Link\"oping University.\\
    E-mail: \{malin.ejdbo, anders.ynnerman, alexander.bock\}@liu.se
  \item
    Joachim Moeyens is with the University of Washington.\\
    E-mail: moeyensj@uw.edu
}

\abstract{\input{0-abstract}}

\keywords{Astrovisualization, planetary defense, uncertainty visualization, asteroids.}

\teaser{
  \centering
  \includegraphics[width=\linewidth]{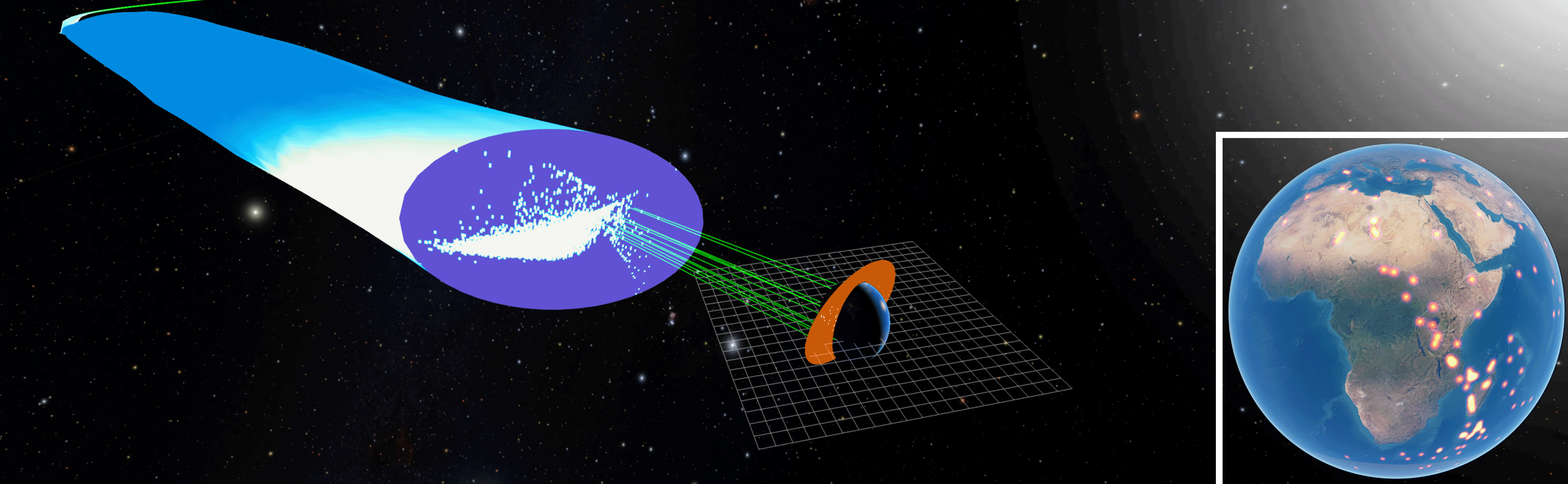}
  \caption{\neoviz~showing a prediction of (99942) Apophis's impact in 2029. The blue \emph{Uncertainty Tube} is a bundled representation of the trajectory samples of which only few (green) impact Earth.  The orange cross-section shows the extent of all impacting trajectories. The reference grid indicates a size of 50,000\,km.  The inset shows the predicted locations for all impacting trajectories on Earth.}
  \label{fig:teaser}
}

\begin{document}
\maketitle

\input{1-introduction}
\input{2-related-work}
\input{3-background}
\input{4-system}
\input{5-case-studies}
\input{6-expert-feedback}
\input{7-conclusion}

\bibliographystyle{abbrv-doi-hyperref}

\bibliography{refs-orbit.bib}

\end{document}

%% file: 0-abstract.tex
We introduce \neoviz, an interactive visualization system designed to assist planetary defense experts in the visual analysis of the movements of near-Earth objects in the Solar System that might prove hazardous to Earth.
Asteroids are often discovered using optical telescopes and their trajectories are calculated from images, resulting in an inherent asymmetric uncertainty in their position and velocity.
Consequently, we typically cannot determine the exact trajectory of an asteroid, and an ensemble of trajectories must be generated to estimate an asteroid's movement over time.
When propagating these ensembles over decades, it is challenging to visualize the varying paths and determine their potential impact on Earth, which could cause catastrophic damage.
\neoviz~equips experts with the necessary tools to effectively analyze the existing catalog of asteroid observations. 
In particular, we present a novel approach for visualizing the 3D uncertainty region through which an asteroid travels, while providing accurate spatial context in relation to system-critical infrastructure such as Earth, the Moon, and artificial satellites.
Furthermore, we use \neoviz~to visualize the divergence of asteroid trajectories, capturing high-variance events in an asteroid's orbital properties.
For potential impactors, we combine the 3D visualization with an uncertainty-aware impact map to illustrate the potential risks to human populations.
\neoviz~was developed with continuous input from members of the planetary defense community through a participatory design process. It is exemplified in three real-world use cases and evaluated via expert feedback interviews.

%% file: 1-introduction.tex
\section{Introduction}
\label{sec:introduction}
Our Solar System is filled with asteroids and comets, making it inevitable that a large object will impact Earth.  An estimated 1 trillion objects orbit the Sun, out of which around 1.3 million asteroids and 4,000 comets have been discovered so far. The number of newly identified small bodies has been growing exponentially in recent years~\cite{demeo2015compositional}. Of particular interest are the \emph{near-Earth objects}~(NEOs). A NEO is a comet or an asteroid whose trajectory brings it close to Earth's orbit around the Sun.
One of the most significant NEOs in history is the \emph{Chicxulub} impactor, an asteroid with an approximate diameter of 10\,km. It struck Earth about 66 million years ago, ending the age of the dinosaurs, marking one of the most devastating events in the history of life on Earth~\cite{SchulteAlegretArenillas2010}. Currently, we know of roughly 75 asteroids of comparable or greater size that would have civilization-ending consequences if they were to collide with Earth.  More recently, the \emph{Tunguska} event in 1908 flattened around 2,000\,$\textrm{km}^{2}$ of forest and was caused by an asteroid estimated to be 50-60\,m~\cite{davydov2021tunguska}.  In 2013, the 20\,m \emph{Chelyabinsk} meteor exploded over Chelyabinsk, injuring over 1,600 people. 

Planetary defense is concerned with detecting potential impactors and devising strategies for their mitigation. The risks posed by these potential impactors are particularly prevalent if the asteroids airburst in or near densely populated areas, which can lead to catastrophic loss of life.
We develop \neoviz~as a visualization system that enables uncertainty-driven analysis of asteroid trajectories using a participatory design process that involves collaboration with planetary defense experts from the B612 Foundation. The foundation's mission is to discover and track potentially hazardous NEOs, raise awareness of their risks, and devise strategies to mitigate their impact.

When observing an asteroid, it is much easier to infer its $(x, y)$ position from a telescope image than the depth in the line-of-sight direction $z$. This introduces inherent asymmetric uncertainty into the orbit fitting and propagation process, which is challenging to visualize. The limited number of observations, combined with the fact that most observations can only occur at night and can take place over several months, leads to a scarcity of data when accurately reconstructing the trajectory of an asteroid, introducing another source of uncertainty. While the quantification of such uncertainty using, for example, covariance matrices, at specific time points and corresponding mitigation strategies has been well-researched~\cite{Yeomans2009DeflectingHazardousObject, RumpfLewisHugh2017}, the visualization of its progression over time and its impact on Earth remains largely unexplored. An effective and accessible tool for communicating the complex, time-varying 3D uncertainty could be invaluable not only for scientific research but also for public outreach and education.

Designing an informative uncertainty representation for asteroid orbit trajectories presents several challenges, including visual clutter, time-varying uncertainty, and significant variability in both time and space. Since the impact probabilities for any given asteroid are fairly low (for example 2.7\% for (99942) Apophis), we can only obtain a representative ensemble of an asteroid's potential trajectories by drawing a large number of samples from its uncertainty distribution. This process creates inevitable visual clutter and heavy computational cost, as many thousands of orbits must be displayed, with only a few trajectories impacting Earth. Additionally, as the uncertainty evolves over time, it is crucial to visualize its progression instead of a collection of static snapshots, which requires a smooth transition between consecutive time steps. Finally, there is large variability in both the magnitude of the uncertainty and the corresponding time period as some uncertainty propagation could span several decades, while others only involve a few hours of data. For some asteroids, their orbit uncertainty could exceed the diameter of the Earth, while for others, their impact point can be predicted accurately. Therefore, we strive to design our visualization system to accommodate and adapt to these wide ranges of scenarios.

In response to these challenges, we developed \neoviz, a visualization system that supports the analysis of individual asteroid trajectory uncertainties. \neoviz~is integrated into the astrovisualization engine \emph{OpenSpace}~\cite{vis19-bock-openspace-system}, which provides the necessary spatial context for asteroid observations. The resulting visualization system offers capabilities surpassing currently available methods, enabling experts to gain important scientific and operational insights into an asteroid's potential movements. The major components of \neoviz~include:

\begin{itemize}[noitemsep]
  \item Given an ensemble of an asteroid's trajectories, we present the \emph{Uncertainty Tube}, a spatiotemporal representation of the volume encompassing all sampled asteroid trajectories. It provides a time-varying visualization of the physical variability of the ensemble.  
  \item At the cross-sections of the Uncertainty Tube, \emph{Interactive cut-planes} shows the sample distribution at particular time steps. For each time step, the user can browse the cut-planes and examine each distribution in detail along the \emph{Uncertainty Tube}.
  \item The \emph{Impact Map} visualizes the likelihood of impact locations on Earth together with an estimated risk analysis for each location.    
\end{itemize}

\noindent The rest of the paper is structured as follows. \cref{sec:relatedwork} discusses prior work on ensemble and uncertainty visualization focusing on curve-based and 3D uncertainties, and then surveys NEO-related visualization tools. \cref{sec:background} provides technical knowledge about the unique characteristics of this uncertain data. We describe our system components in detail in \cref{sec:system} and three case studies in \cref{sec:case-studies}. Finally, we summarize the results of four qualitative interviews with planetary defense experts in \cref{sec:expert-feedback} and discuss limitations and future work in \cref{sec:conclusion}.

%% file: 2-related-work.tex
\section{Related Work}  \label{sec:relatedwork}
Uncertainty quantification is a vibrant field of study in astronomy due to the inherent error margins introduced in the data acquisition process.  Thus far, visualization has, however, only rarely been used to convey this uncertainty in 3D, because of the data complexity and the large variability in both time and space. In this section, we first review relevant prior works on ensemble and uncertainty visualization. We then discuss existing methods for analyzing the trajectories of NEO.

\subsection{Ensemble and Uncertainty Visualization}
Significant advances have been made to ensemble visualization due to the increased availability of data, which is often a result of running multiple instances of the same physical simulation~\cite{WangHazarikaLi2019}. Uncertainty visualization is a common technique to effectively communicate the corresponding uncertainty. Several surveys detailing its development and applications are available~\cite{PangWittenbrinkLodha1997, Pang2008, PotterRosenJohnson2012, KamalDhakalJavaid2021}.  Recently, Wang~\etal~\cite{WangHazarikaLi2019} surveyed visual analysis techniques on ensemble simulation data and partitioned scientific visualization techniques of spatial data into four categories: point-oriented, curve-oriented, surface-oriented, and volume-oriented approaches.  Highly relevant to our work, we next discuss curve-based uncertainty techniques, especially in the broader context of 3D and temporal uncertainties; an area which in itself has seen limited attention from the research community thus far.

Zhang~\etal~\cite{ZhangChenLi2021} used variable spatial spreading and spaghetti plots to show the uncertainty in the attribute variable dimension. Their visualization was used to identify features of interest and perform comparative analysis in a weather forecast ensemble. Inspired by conventional boxplots, Whitaker~\etal~\cite{WhitakerMirzargarKirby2013} introduced contour boxplots by computing order statistics on an ensemble of contours. The contour boxplot was then further enhanced through the use of curve boxplots~\cite{MirzargarWhitakerKirby2014} as an extension to streamlines and pathlines.  Liu~\etal~\cite{LiuPadillaCreemRegehr2019} moved away from traditional error cones and proposed an implicit uncertainty visualization by selectively sampling from the original ensemble and reconstructing a spatially well-organized ensemble. They applied their method to tropical cyclone forecast tracks and demonstrated that their visualization could assist scientists in the exploration and estimation of storm damages while preventing visual confusion. Analogous to their approach, we employ a combination of a spatially distributed ensemble, an extension of an error cone into a time-varying 3D structure, and the visualization of individual trajectories to more effectively convey the potential hazards associated with a particular near-Earth object.

Visualizing uncertainties in 3D is known to be challenging. Guo~\etal~\cite{GuoYuanHuang2013} developed \emph{eFLAA} to characterize the variations among 3D ensemble flow fields.  They incorporated a 3D navigation and a timeline view to provide an overview and regional comparison capabilities based on the user's selection. A 3D uncertainty band was also used to visualize geospatial uncertainties. Fersl~\etal~\cite{FerstlBurgerWestermann2016}, on the other hand, visualized uncertainty in vector field ensembles using confidence lobes, which are constructed using dimensionality reduction and clustering on the ensemble.
To prevent disastrous underground utility strikes, Li\etal~\cite{Li2015UncertaintyAwareGeospatial} used multipatch surface models to construct 3D probabilistic uncertainty bands that enclose the true utility lines while accounting for positional uncertainties. These uncertainty representations inspired the design of the \emph{Uncertainty Tube} in \neoviz.

Zhang~\etal~\cite{ZhangDemiralpLaidlaw2003} used streamtubes to visualize 3D diffusion tensor MRI data, where streamlines are grouped to generate a tube representation. The main differences to our approach are that we must consider all sampled trajectories due to the asteroids' small impact probabilities, whereas it was sufficient to characterize the main behavior of the streamlines in their data by selecting a small subset of trajectories; we need to perform time stitching and ensure smooth transition during the time steps while their method shows a static tube visualization.

In astrophysics, few visualizations exist to facilitate exploration of uncertainty.  Bock~\etal~\cite{BockPembrokeMays2015} proposed a system to study the uncertainty of space weather simulations. They provided a multi-view visualization providing comparative analysis capabilities, the exploration of time-varying components in the data, and a volume rendering of the simulations.  In a recent survey on the visualization of astrophysics, Lan~\etal~\cite{LanYoungAnderson2021} identified uncertainty visualization as a significant challenge in the field, offering numerous research opportunities.  In this work, we present a system that represents an effort to address this gap.

\subsection{Visualization in NEO Risk Analysis}
The analysis of NEO risk is a well-explored subject with considerable effort dedicated to quantifying the implications of asteroid orbit uncertainty.  However, visualization techniques in the field have remained rudimentary. Most studies use bar charts, line plots, and scatter plots as the main methods of communicating their scientific insights.

Yeomans~\etal~\cite{Yeomans2009DeflectingHazardousObject} were among the first to raise concerns about (99942) Apophis, discussing keyhole events in both 2029 and 2036, and considering potential deflection strategies for near-Earth asteroids.  They introduced a compelling ``keyhole map'', an adaptation of a line plot, that is generated by systematically probing the current uncertainty region of the trajectory and propagating each potential trajectory forward.  The map offers an overview of how variations in the current orbit could result in catastrophic consequences decades later.  Paek~\etal\ then further elaborated on the existence of multiple keyhole events for Apophis in 2036 and investigated various deflection strategies~\cite{Paek2016AsteroidDeflectionCampaign, Paek20202OptimizationDecision-making}.

Wheeler~\etal~\cite{Wheeler2017SensitivityAsteroidImpact} analyzed which asteroid properties and entry parameters cause the most damage in case of an impact. They used tornado plots and histograms to conduct a comparative analysis of the risk uncertainties for four nominal asteroid sizes ranging from 50--500\,m in diameter and concluded that the impact location contributes most to impact risk uncertainty, followed by its size and velocity.

Similar to our goal of facilitating asteroid risk analysis, Rumpf \etal~\cite{Rumpf2019ProjectingAsteroidImpact} proposed a probabilistic visual representation of impact corridors. They also incorporated uncertainty information by using a Monte Carlo method to sample the orbital solution state space based on the covariance matrix and propagating the orbits forward to compute impact locations. They visualized the spatial impact probability distribution of each impact location using a 2D scalar field and scatter plots. In a later work, Rumpf~\etal~\cite{Rumpf2020DeflectionDrivenEvolution} investigated how uncertainty and risk were affected by a planned deflection of an uncertain asteroid. In their work, they visualized the impact corridor with a 2D scatter plot. Norlund~\etal~\cite{NEOMiSS} provided a more reliable prediction of human casualties and infrastructure risks caused by a NEO impact in \emph{NEOMiSS}. They developed a behavior-based evacuation model by combining the physical effects of a potential impact, the historical insights into the impact region such as natural hazards and local building properties, and crowd-sourced information regarding transportation infrastructure. They overlaid a 2D map with the population density of the impact region and color-coded each area based on the predicted human casualties. The \emph{Impact Map} in \neoviz~is inspired by this approach. However, we extend it to first show the location within its 3D context, providing the ability to show other georeferenced maps simultaneously.

These state-of-the-art visualizations in planetary defense demonstrate the usage of effective visualizations on their own. However, there is currently a severe lack of a contextualized integration of these disparate techniques.  The ability to investigate the underlying uncertainty of impact locations strongly depends on the ability to inspect the NEO trajectories in a time-varying 3D space to gain the necessary insights.  \neoviz~aims to address this by combining improved versions of several visualization techniques into a common reference frame and simultaneously supporting interactive exploration.

%% file: 3-background.tex
\section{Background}  \label{sec:background}
\begin{figure*}[t]
  \centering
  \includegraphics[width=\linewidth]{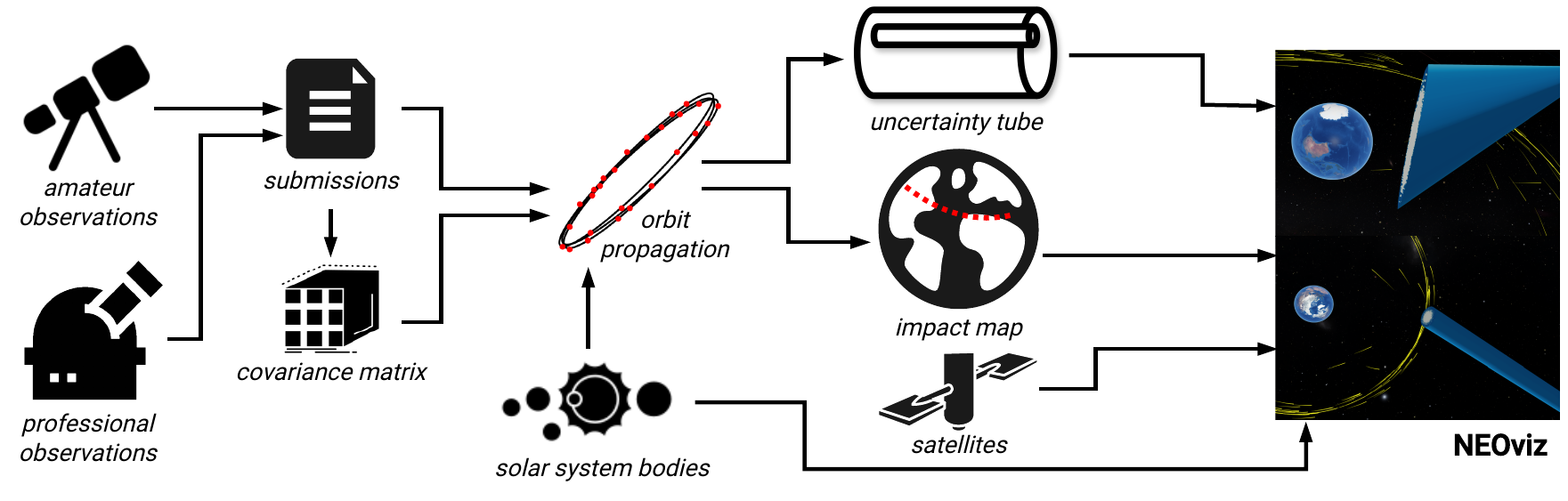}
  \caption{A systematic overview of \neoviz. NEO observations and their associated uncertainties are used to propagate trajectories into the future. These trajectories are then visualized as a collection using the \emph{Uncertainty Tube}s and an \emph{Impact Map} in the case of an impact. The results are presented in context to other solar system bodies and artificial satellites.}
  \vspace*{-\baselineskip}
  \label{fig:system}
\end{figure*}

This section describes the domain background focusing on the characteristic nature of NEO orbits and the uncertainty associated with the data acquisition and orbit propagation.  This section also describes how the input data for \neoviz~was created from these initial observations.

While humanity's ability to discover asteroids has drastically improved, there is still much left to do. The vast majority of asteroid discoveries have so far been contributed by ground-based, optical surveys.  Starting in 2025, the Vera C. Rubin Observatory will begin a 10-year survey called the Legacy Survey of Space and Time~\cite{Ivezic2019}, which is estimated to contribute over 5 million new asteroid discoveries, nearly quadrupling the known asteroid population. This increase in data exacerbates the need for a visual analytical tool to systematically examine the new discoveries~\cite{Ivezic2019, Jones2018}.  Additionally, space-based telescopes not only serve to characterize the sizes of asteroids but can also discover asteroids that are not visible from the ground.

Most asteroids and comets are located in the Main Asteroid Belt between Mars and Jupiter or the Kuiper belt beyond Neptune.  Over short time spans, typically a few weeks, their motion on perturbed Keplerian orbits can be approximated with simple 2-body dynamics due to the Sun's overwhelming gravitational influence.  However, over longer time spans, the gravitational influence of the planets and other massive objects needs to be accounted for to accurately predict the positions of asteroids.  In addition to gravitational effects, such as collisions in the Main Belt or close encounters with a planet, asteroids and comets may also be subject to non-gravitational forces such as outgassing, the Yarkovsky effect~\cite{Vokrouhlicky2000} that introduce high levels of uncertainty and make long-term predictions challenging.

Asteroids are typically discovered at the magnitude limit of astronomical surveys and near opposition where they appear brightest.  The orbital uncertainty within the arc of observations is well-approximated by a 3D ellipsoid, where the uncertainty of the orbit derived from observations is commonly represented as a 6D covariance matrix. The six dimensions correspond to the orbital parameters of an asteroid, namely its Cartesian coordinates $(x, y, z)$ and the velocity in each direction $(v_x, v_y, v_z)$. The diagonal of the covariance matrix depicts the variance of each orbital parameter. Beyond the arc of observations, the uncertainty region instead becomes extended or elongated (referred to as a ``banana-oid'').  This elongation occurs due to Keplerian shear, in which orbits closer to the Sun will possess greater velocities than orbits further from the Sun.  Propagated over time, the variation in distance from the Sun, and by extension, the variation in velocity, causes the uncertainty region to become elongated or extended.

Further observations through observation campaigns or via precovery~\cite{boattini2001arcetri} in existing archives refine the orbital parameters of the object and influence its uncertainty.  Impact probability studies typically involve sampling the covariance matrix using Monte Carlo methods for variant orbits and propagating these forward.  The ratio between impacting orbits over the total number of variants is then reported as the object's impact probability.  For this sampling technique, since each variant is individually propagated, Monte Carlo sampling is both the most accurate but also the most computationally expensive to perform. 

A global effort, coordinated in part by the United Nations Office for Outer Space Affairs, NASA, ESA, JAXA, and others, aims to identify and mitigate any potential future asteroid impacts.  The Asteroid Institute, a program of the B612 Foundation, is a research institute that participates in the planetary defense effort by building the tools to process the next generation of astronomical data.  One of these tools is the Asteroid Discovery Analysis and Mapping platform (ADAM), which is designed to enable compute-intensive research in astronomy and currently has services to perform precovery searches for asteroid observations, asteroid discovery searches in telescopic data, and Monte Carlo impact probability calculations.  Critical to the planetary defense effort is the ability to communicate the nature of the impact hazard to both the planetary defense community and to the general public. The visualization of orbital uncertainty regions and risk corridors are necessary tools to facilitate such communication.

For each of the asteroids detailed in the case studies in \cref{sec:case-studies}, we used \mpcq\footnote{\url{https://github.com/B612-Asteroid-Institute/mpcq}~} to gather observation and submission histories.  \mpcq~ allows the user to query the Small Bodies Node replica of the Minor Planet Center (MPC) database for observations of known asteroids and details about their submission history. As observatories conduct their campaigns, bundles of observations are submitted to the MPC typically at the end of a night of observation. The timestamp associated with each submission is often defined as the timestamp of the last observation within the submission.  Once a submission of observations is accepted by the MPC, the orbits of any known objects are updated.

%% file: 4-system.tex
\section{System}  \label{sec:system}
This section outlines the design for our orbit uncertainty visualization system.  We begin by detailing the \emph{Uncertainty Tube} generation process, followed by a discussion of the graphical rendering techniques. To facilitate an efficient scientific discovery process, we also implement several interactivity features. Lastly, for potential impactors, we create an \emph{Impact Map} that displays the predicted impact locations along with associated risk probabilities.

To determine an orbit from a submission of observations, we use the software \findorb\footnote{\url{https://www.projectpluto.com/find_orb.htm}~}, which is extensively used by the planet defense community~\cite{Larson2003}.  Given a set of observations from a submission, \findorb~determines the best-fit orbit, whose uncertainty is calculated as a covariance matrix during the least-squares orbit correction optimization. The orbit fitting process for each object is run iteratively over time, with every new submission including the observations from previous submissions. The outcomes of the orbit determination process include the observations, submission history, best-fit orbit derived from these observations, and the time-varying uncertainty.

In addition to \findorb, we use the {\adamcore}\footnote{\url{https://github.com/B612-Asteroid-Institute/adam_core}~} library for various orbit-related computations. {\adamcore}~contains the shared set of utilities used by the services hosted on the ADAM platform and also provides access to orbital integrators such as PYOORB~\cite{Granvik2009}.
In \neoviz, we employ {\adamcore} to propagate orbits forward in time and perform Monte-Carlo sampling using the covariance matrix.

\subsection{Uncertainty Tube Generation} \label{sec:tube-generation}
For each NEO, the submission data consists of a batch of timestamped observations.  For each submission time $t$, we obtain a covariance matrix that represents the uncertainty of the orbit from orbit fitting.  The uncertainty is time-varying and asymmetrical, which means that the uncertainty values can differ greatly for each direction over time.  The goal is to encapsulate the uncertainty using ellipse slices at each submission time that are oriented perpendicular to the mean velocity of the samples.  Connecting all ellipse slices results in a 3D tube and enables the expert to visually inspect it within its spatial context.  To construct the \emph{Uncertainty Tube}, we introduce a three-step process:
\begin{enumerate}[noitemsep]
  \item \emph{Orbit propagation and variant sampling}: Compute the best-fit orbit using the submissions and their associated uncertainties and sample orbit variants from the resulting covariance matrix.
  \item \emph{Ellipse slice computation}: For each time step, compute an ellipse slice perpendicular to the mean velocity of the samples to approximate the uncertainty region of the orbit.
  \item \emph{Time stitching}: Find correspondences between neighboring ellipse slices by sampling a fixed number of points on the perimeter of each ellipse in a deterministic order. Connect the ellipse slices into a 3D tube.
\end{enumerate}

An alternative approach to using ellipses could be to use convex hulls, which can provide a more accurate approximation of the uncertainty boundary. However, establishing correspondences between convex hulls during time stitching presents significant challenges, and the resulting 3D structure during interpolation might not be meaningful and be more difficult to interpret. We also considered constructing the ellipses through a 2D Eigendecomposition of the covariance matrix, but such ellipses may only encompass the majority of the trajectories, which is insufficient for our purpose.

\begin{figure}[b!]
  \centering
  \vspace{-6mm}
  \includegraphics[width=0.8\columnwidth]{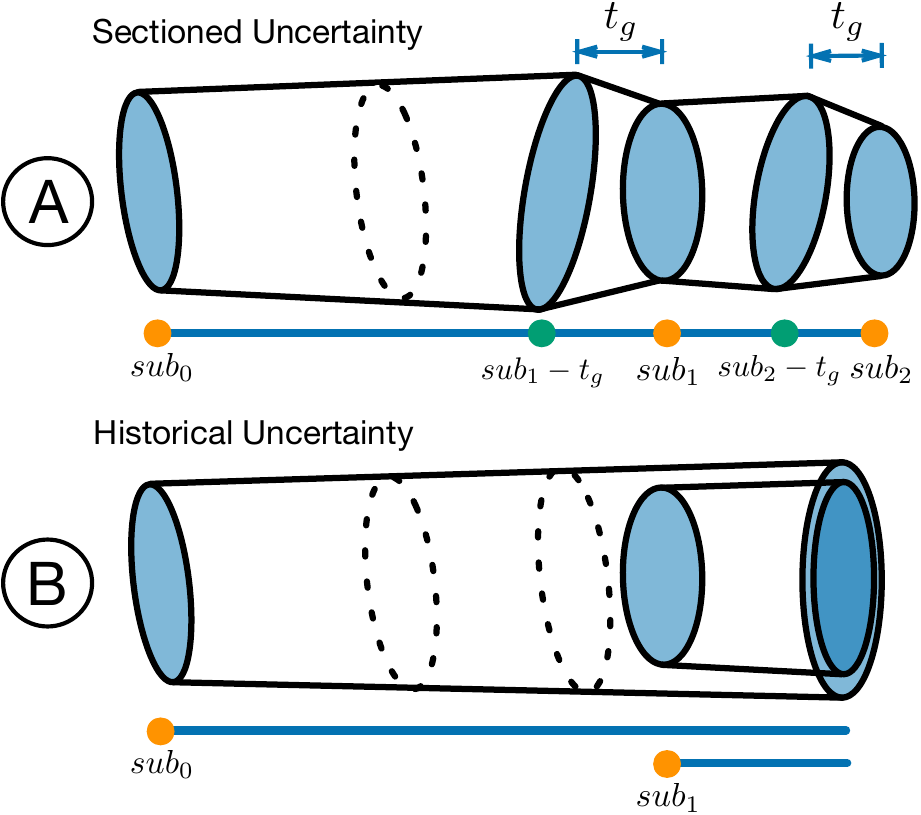}
  \caption{Illustrations of the two different \emph{Uncertainty Tube} representations: (A) \emph{sectioned uncertainty} and (B) \emph{historical uncertainty}. The timeline below each image shows the time ranges for each \emph{Uncertainty Tube}.}
  \vspace*{-\baselineskip}
  \label{fig:system-data}
\end{figure}

\begin{figure*}[t]
  \centering
  \includegraphics[width=\linewidth]{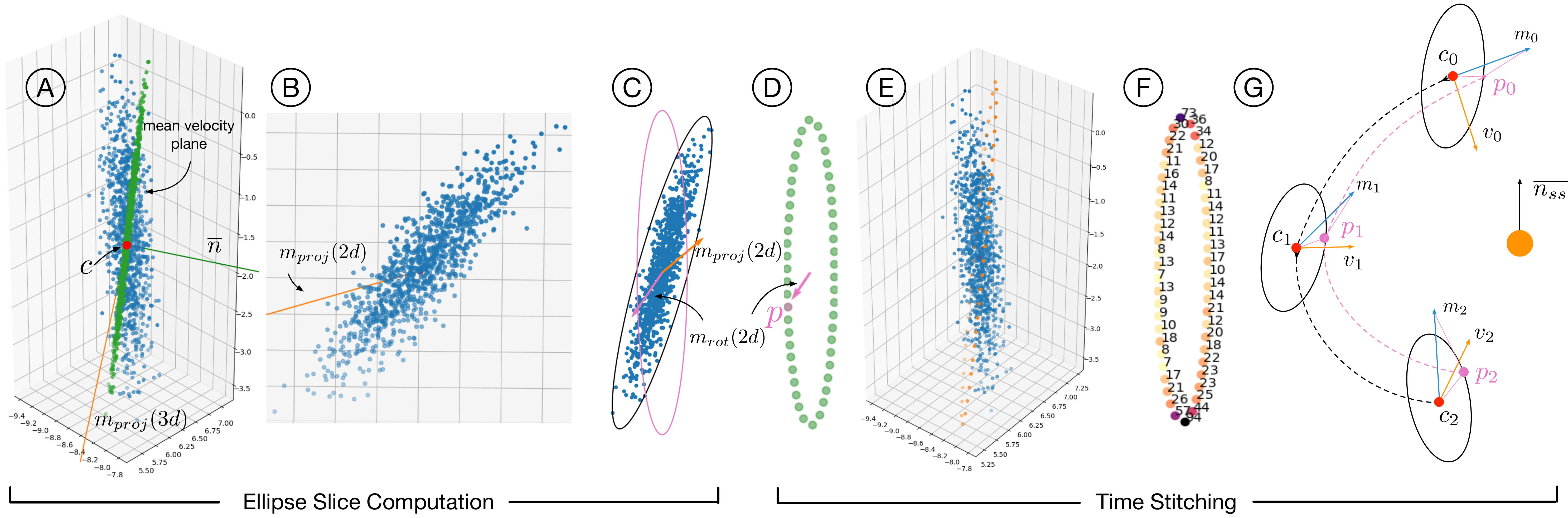}
  \caption{Ellipse slice computation and time stitching pipeline. At each time step, the trajectories are projected into a 2D space to calculate a minimum enclosing ellipse, which is used as an uncertainty representation. A consistent reference frame that is co-rotating with the NEO orbits is then used to connect adjacent ellipses and generate the \emph{Uncertainty Tube}.}
  \vspace*{-\baselineskip}
  \label{fig:ellipse-comp}
\end{figure*}

\para{Orbit propagation and variants sampling}
Given the entire submission history of an object, we utilize only the observations available up to a given time.  When an asteroid is first discovered with only a few observations in the submission history, \findorb~fails to identify an orbit.  Therefore, we must determine the initial submission time at which a sufficient number of observations are available for \findorb~to identify an orbit. Proceeding forward from this point in time, we can freely select a start and end time for the orbit propagation.  For each subsequent submission time, we find a new best-fit orbit, including the covariance matrix, considering all past observations.

We introduce two uncertainty representations for the 3D tube: \emph{sectioned uncertainty} and \emph{historical uncertainty}.  \emph{Sectioned uncertainty} applies to events where all submission data is present, whereas \emph{historical uncertainty} is designed for predictions of asteroid trajectories. For the \emph{sectioned uncertainty} representation, as illustrated in~\cref{fig:system-data} (A), we start at a submission $sub_0$.  We then propagate the associated best-fit orbit forward to shortly before the next submission time $sub_1$.  We call all such orbits the \emph{propagated best-fit orbit}.  We enforce a time gap ($t_g$ in \cref{fig:system-data}) between the end of $sub_0$ orbit and the start of the $sub_1$ orbit to ensure that the changes in size and direction of the tube between submissions can be clearly seen without introducing instantaneous changes that would draw unwanted attention.  When reaching the last submission, we propagate the orbit forward to a user-specified end time, at which point the covariance matrix for this propagated best-fit orbit is reconstructed using a Monte Carlo sampling method (see~\cref{sec:background}).  We then uniformly sample a large number of orbit variants from this distribution in the same time frame, \ie~from submission times $sub_0$ to shortly before $sub_1$.  Using this procedure, we produce a large sample of possible orbits that represent the orbit uncertainty at each submission time until we reach the next batch of observations.  The \emph{Uncertainty Tube} would be sectioned by the submissions, where the shape of the tube is recomputed and corrected at each new submission time.

For the \emph{historical uncertainty} representation, we also start at submission time $sub_0$.  However, instead of correcting the propagation with each subsequent submission, we consider only the observations up to $sub_0$ for the orbit propagation.  As illustrated in \cref{fig:system-data} (B), we propagate forward from $sub_0$ to a user-specified end time without considering new submissions.  Using this orbit and covariance matrix, we repeat the same procedure as in the \emph{sectioned uncertainty} scenario and uniformly sample a large number of orbit variants.  At $sub_1$, we repeat the process and use the new best-fit orbit to obtain a new set of orbit variants and create another \emph{Uncertainty Tube}.  The \emph{historical uncertainty} representation recreates historical events to investigate the predictions made using only the observations available at a particular point in time.

\para{Ellipse slice computation}
To capture the orbit's uncertainty structure in 3D, we generate ellipse slices that reflect the shape of the orbital distribution at various time steps.  We use an adaptive time sampling strategy based on the time interval between two consecutive submissions that produces one ellipse slice per submission or at least one slice per day. A clear change in the \emph{Uncertainty Tube} at each new submission is indicated by the time gap, denoted as $t_g$ in \cref{fig:system-data}, defined as a small percentage of the corresponding submission interval. For the \emph{Sectioned Uncertainty Tube} illustrated in \cref{fig:system-data} (A), since the submission interval between $sub_0$ and the end of $sub_0$, namely $sub_1 - t_g$, is larger than one day, we insert an additional ellipse slice in between the starting and ending ellipse slices. In the case of the \emph{Historical Uncertainty Tube} (B), ellipse slices continue to be added as time progresses forward at an interval of approximately one day. In particular instances where an expert seeks to closely inspect a certain time period, the temporal resolution is further increased to produce a more refined \emph{Uncertainty Tube} section, following the ``Overview first, details on demand'' mantra~\cite{shneiderman2003eyes}. From this process, we obtain a predefined array of time steps.

\cref{fig:ellipse-comp} (left) provides an overview of our ellipse slice computation pipeline.
At a given time step, the positions of all sampled trajectories form a point cloud, illustrated in blue in \cref{fig:ellipse-comp} (A).
To represent the uncertainty, the aim is to compute a 2D ellipse slice for this 3D point cloud, reducing the problem to a 2D task of finding the minimum enclosing ellipse of the point cloud in a plane.

We define the center $c$ of the point cloud to be the center of the 3D minimum volume enclosing ellipsoid. Using the point cloud's center $c$, marked in red in \cref{fig:ellipse-comp} (A), and the mean velocity direction of the variants as the normal vector $\overline{n}$, we determine a 2D \emph{mean velocity plane}.  All points in the 3D point cloud are then projected onto the plane in the direction of their instantaneous velocity vector. In \cref{fig:ellipse-comp} (A), the blue points are the original 3D point cloud and the projected points are drawn in green.  These projected points are then transformed from the \emph{mean velocity plane} into the x-y plane shown in \cref{fig:ellipse-comp} (B), where it is now a 2D problem of finding the minimum area enclosing ellipse for the projected 2D point cloud.  The representative ellipse is found using the algorithm by Bowman and Heath~\cite{BowmanHeath2023}. \cref{fig:ellipse-comp} (C) shows an example ellipse of the 2D point cloud in black. The resulting ellipse, together with the distribution of the 2D point cloud, is transformed into a cut-plane at the corresponding time point on the \emph{Uncertainty Tube}.

\para{Time stitching}
We compute an ellipse slice for each of the predefined time steps. We connect ellipses from adjacent time steps by establishing correspondences between them. The goal is to uniformly sample the same number of points from the perimeter of each ellipse and match the points between any two adjacent slices. This way we prevent the \emph{Uncertainty Tube} from twisting during interpolation over time.

In order to establish correspondences, we first need to determine the position on each ellipse to start the sampling process. We need to make sure that the starting point is approximately the same for each ellipse with respect to the center of the ellipse and the location of the sun. We rely on two vectors that are relatively constant with respect to the asteroid, the normal of the fundamental plane of the Solar System $\overline{n_{ss}}$ and the direction from the center of the 3D point cloud $c_i$ to the Sun, denoted $v_i$ for each time step $t_i$, see \cref{fig:ellipse-comp} (G).  Both the normal and the directional vectors are computed using NASA's SPICE library~\cite{Acton1996Spice, Acton2018Spice}.  We take the cross product of these two vectors $v_i$ and  $\overline{n_{ss}}$ and obtain the vector $m_i$. However, $m_i$ is often not on the mean velocity plane.  We thus perform an orthogonal projection of $m$ onto the mean velocity plane, denoted $m_{proj}(3d)$ in \cref{fig:ellipse-comp} (A).  As we transform the points to 2D, we also transform the vector onto the 2D plane, shown as $m_{proj}(2d)$ in \cref{fig:ellipse-comp} (B).  We rotate the ellipse for the major and minor axes to align with the x and y axes and the rotated $m_{proj}(2d)$, presented as $m_{rot}(2d)$, is drawn in pink in \cref{fig:ellipse-comp} (C) and (D).  For each ellipse, the sampling starts at intersection point $p$ of $m_{rot}(2d)$ and the ellipse and uniformly continues for a fixed number of points in a clockwise direction to result in the \emph{sampled points} on the boundary of the ellipse. \cref{fig:ellipse-comp} (E) shows the sampled points (orange) in relation to the 3D point cloud (blue). Finally, we interpolate between corresponding points from adjacent ellipses, e.g. between $p_0$ and $p_1$, $p_1$ and $p_2$ in \cref{fig:ellipse-comp} (G).

We encode further the density distribution of the orbit variants as the surface color of the tube in the \emph{Uncertainty Tube} rendering stage. This technique was desired by the domain expert to be able to distinguish distribution changes over long time periods at a quick glance and to see whether they are equally distributed or skewed (see~\cref{fig:teaser}, left). For an ellipse slice, we compute the density information by constructing a ray from the center of the ellipse passing through each point, and computing the intersection between the ray and the ellipse.  Then, for each sampled point on the ellipse, we count the number of intersection points surrounding the sampled point within a given radius $r$. We assign this value as the density value of each sampled point on the ellipse. In \cref{fig:ellipse-comp} (F), we color each sampled point based on its associated density value, with the darker colors indicating a higher density. To select the appropriate $r$, we take the minimum of two quantities: the length of the minor axis of the ellipse, and the distance between two adjacent sample points on the ellipse's boundary.

\subsection{Uncertainty Tube Rendering}  \label{sec:tube-rendering}
\begin{figure}[b!]
  \centering
    \includegraphics[width=\linewidth]{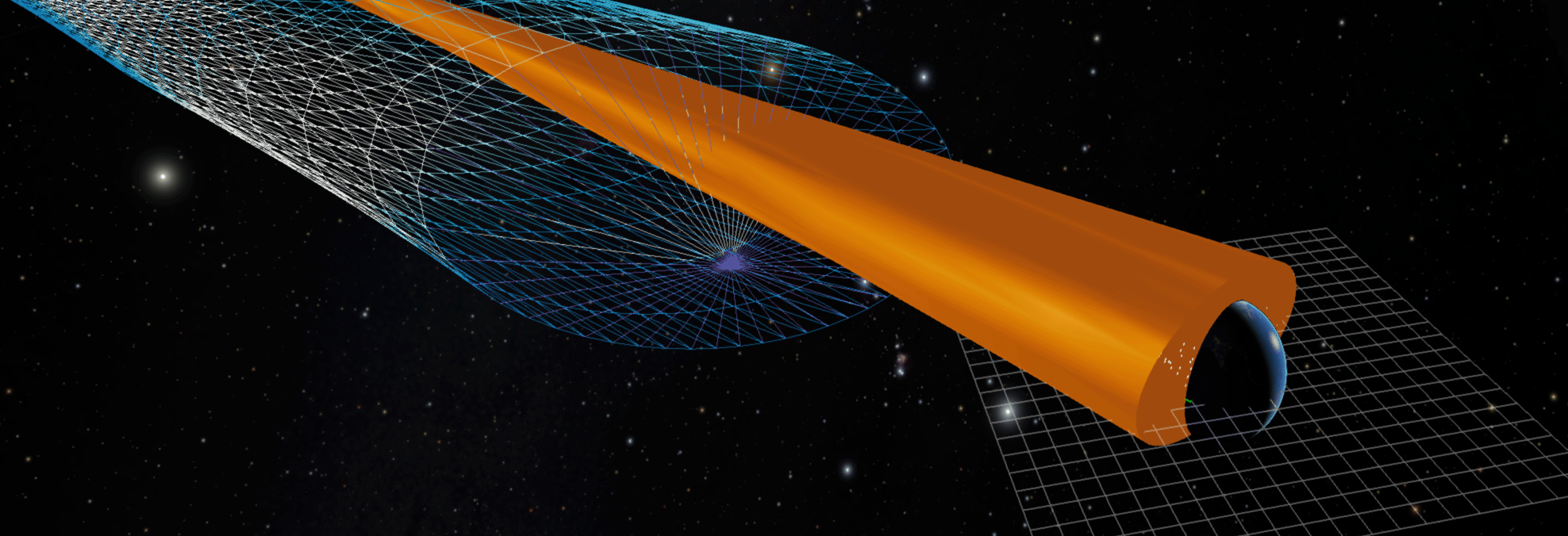}
    \caption{Toggling the \emph{Uncertainty Tube} into a wireframe mode enables the inspection of trajectory variants inside the tube in the case of nested uncertainty. The blue tube shows the uncertainty of all trajectories, the orange tube shows ones intersecting Earth.}
  \label{fig:rendering-tube}
\end{figure}

Using the ellipse data from the previous step, a tube is generated by connecting adjacent slices in a triangle mesh to form a 3D structure that visualizes the known uncertainty of the sampled trajectories (see~\cref{fig:rendering-tube}).  The parameter and the transfer function used to colorize the surface are controlled by the user.  Furthermore, the user can choose to show the ellipse images created in the previous step on the ``lid'' of the tube (\cref{fig:teaser} (left)).  Lastly, the tube can be rendered as a wire-frame which is used to inspect individual trajectories that lie on the inside of the tube, see \cref{fig:rendering-tube} for an example.

The user can change the simulation time in \neoviz~to simultaneously control the position of all objects. The extent of the \emph{Uncertainty Tube} is only shown until the selected simulation time. Since this selected time might lie in between two ellipse slices, to display reasonable information at this time, both the tube's vertices and the cut-plane information are interpolated between the preceding and succeeding time steps using linear interpolation.  As the relative time distance between ellipse slices is fairly small, a linear interpolation was deemed to be sufficient and not to introduce additional errors.

In the case of visualizing the \emph{historical uncertainty}, in which multiple tubes are calculated, the domain expert chooses which \emph{Uncertainty Tube} to show by providing individualized stop times for each representation. This provides the ability to comparatively study the uncertainty's temporal evolution when a specific new submission is added, thus validating its impact on the orbital characteristics.

\subsection{Impact Map}  \label{sec:impact-map}
In addition to the trajectories of the NEOs, it is also vital to show an accurate representation of the location where a potential impactor might collide with the Earth.  As these events are rare, it is beneficial to provide the geospatial context to the impact locations in relation to the 3D trajectory visualization described above.  Traditionally, these maps are represented as \emph{impact corridors} which show a trail on Earth where a NEO might impact.  However, these impact corridors rarely display uncertainties and never include information about the original trajectory of the sample variant that resulted in a specific impact location.

\begin{figure*}[t]
  \centering
  \centering
  \subfloat[Uncertainty magnitude vs number of observations]{\includegraphics[width=0.2475\linewidth]{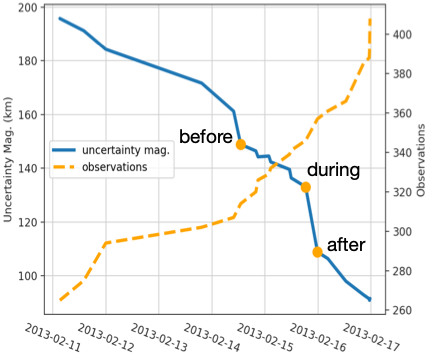}}
  \hfill
  \subfloat[Before close approach]{\includegraphics[width=0.2475\linewidth]{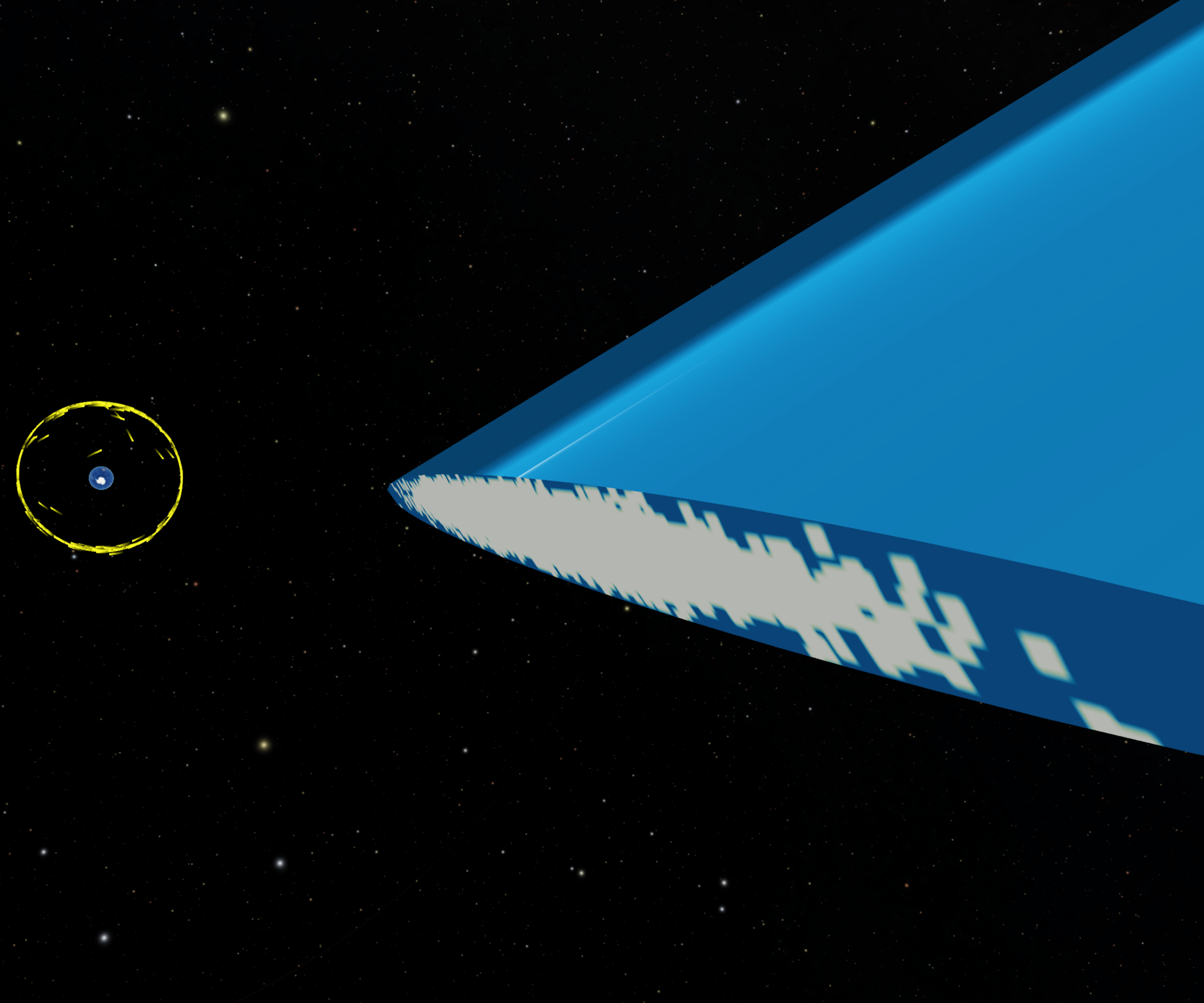}}
  \hfill
  \subfloat[During close approach]{\includegraphics[width=0.2475\linewidth]{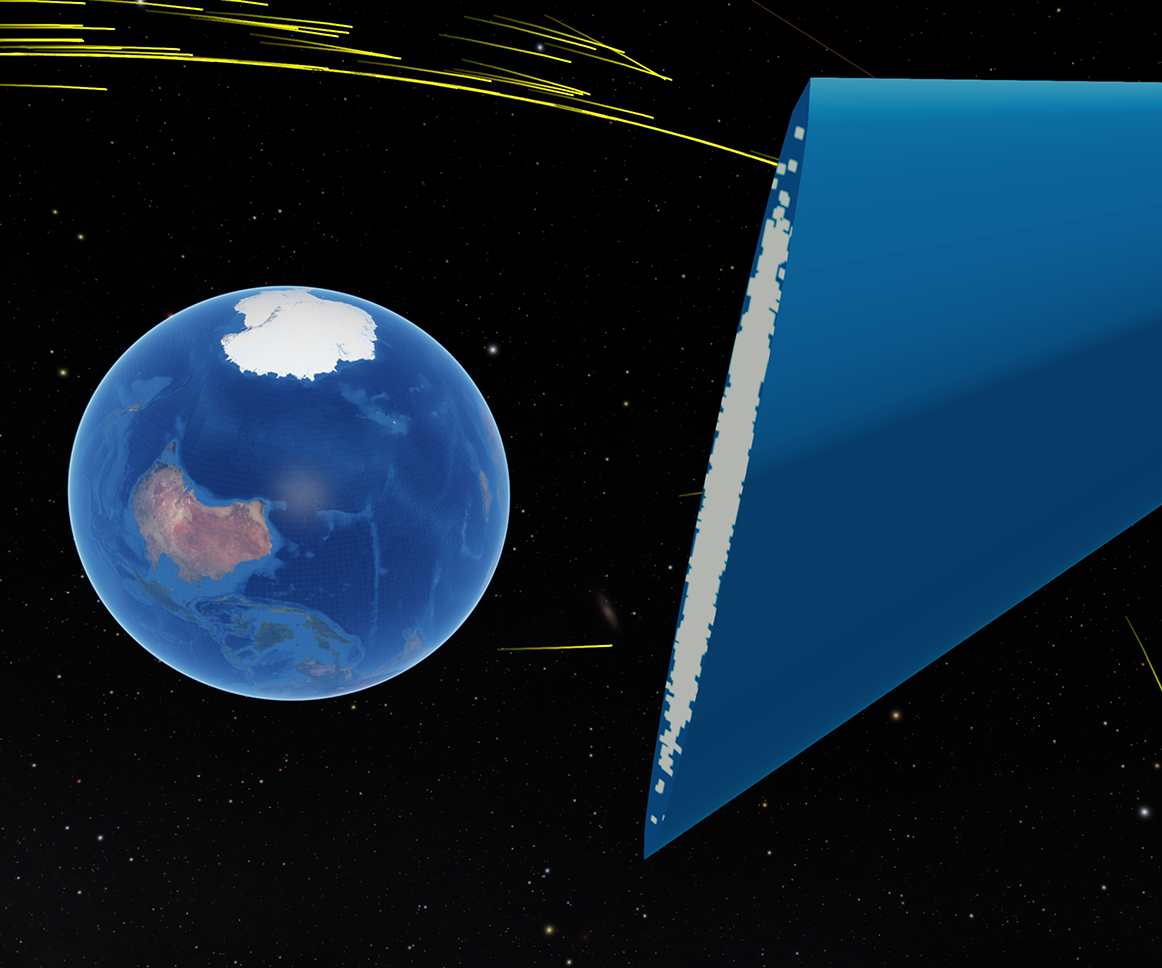}}
  \hfill
  \subfloat[After close approach]{\includegraphics[width=0.2475\linewidth]{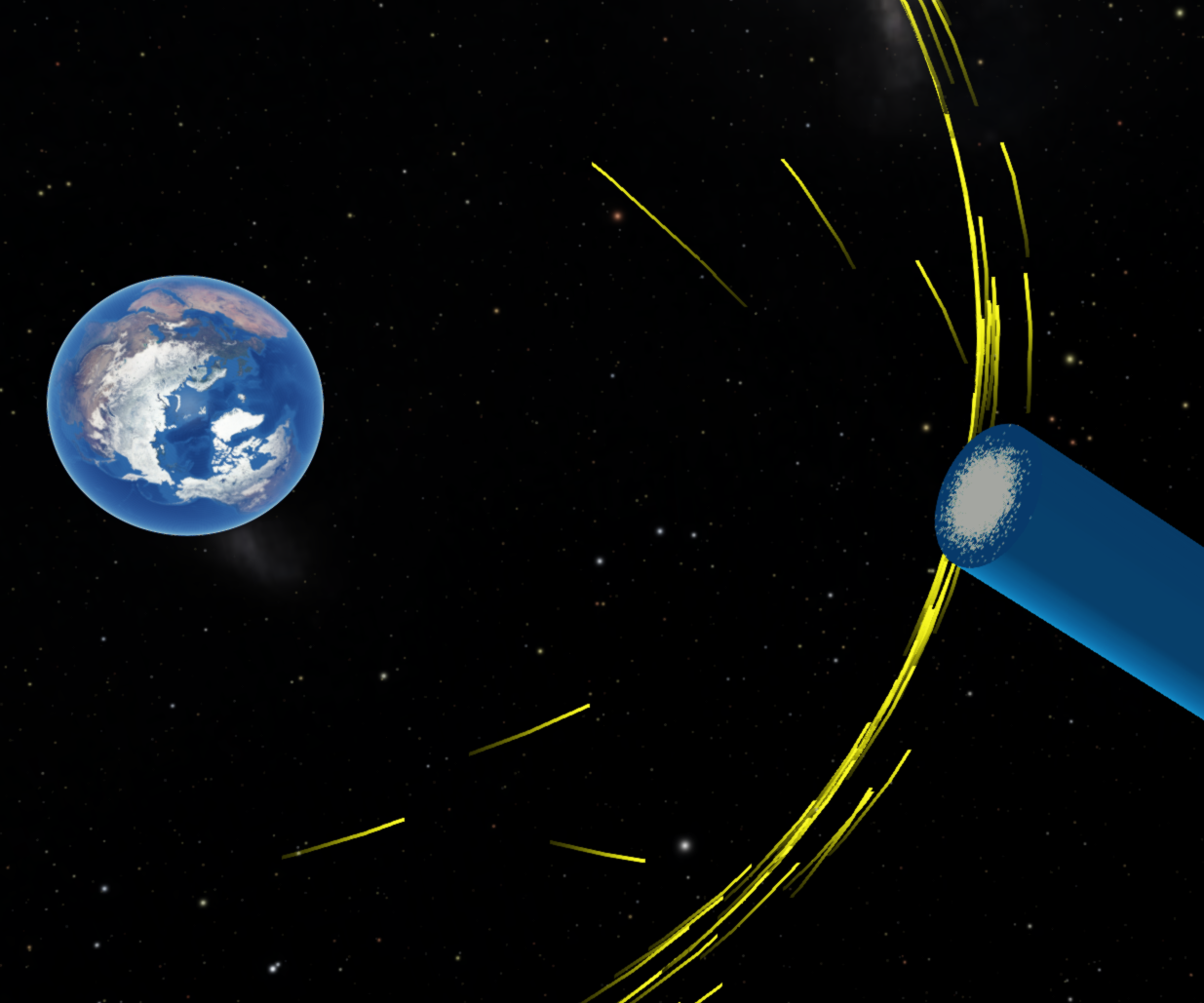}}
  \caption{During asteroid (367943) Duende's close approach on Feb 15\dateth, 2013, the positional uncertainty increases due to the gravitational pull of Earth, while a second source of uncertainty, a lack of observations, decreases.  \neoviz~shows the transition accurately as the uncertainty shows a skew towards the Earth before the closest approach and a more uniform distribution after the encounter.  The orange points in (a) correspond to the times at which figures (b), (c), and (d) were created.  The trails around Earth show the Geostationary satellites at a distance of 36,000\,km.}
  \vspace*{-\baselineskip}
  \label{fig:da14-close}
\end{figure*}

The trajectories calculated in the previous step are also used to calculate the predicted impact location on the surface. After all impact points are collected, these are drawn as circles onto an image using the equirectangular map projection.  The radius of the circle is fixed for each impact, and is adjustable to represent the potential impact energy, if such information is available for a given impactor.  The circle furthermore uses a Gaussian falloff to convey the remaining uncertainty of these impact locations due to atmospheric effects that are not accounted for in the trajectory calculation.  All impact circles are composited using additive blending to result in a final map that represents a higher impact likelihood with a higher numerical value.

The impact probability map (see~\cref{fig:teaser}, right) shows areas with a higher impact probability with a brighter color than areas with individual dots.  This indicates where on Earth it would be more or less likely for the asteroid to impact, giving an approximate of the impact probability for that location.  This map, however, only shows the \emph{impact probability} irrespective of what is located underneath the impact location.  To visualize the risk to human lives, a second map is generated that highlights areas with large human populations.  This map is generated by combining the impact probability map with the location of human-made artificial lighting on Earth, which has been shown to accurately represent the human footprint on Earth~\cite{kyba2015high}. Similar to the impact probability, this map is colorized using a transfer function that maximizes the data's visibility when projected onto the globe in the 3D visualization and when composited with existing satellite images.

%% file: 5-case-studies.tex
\section{Case Studies}  \label{sec:case-studies}
The following three case studies demonstrate various use cases of \neoviz~to facilitate the scientific discovery process of planetary defense experts.  We investigate three asteroids of interest: (367943) Duende, (99942) Apophis, and \designation{2023}{CX}{1}. \edit{For each asteroid, we generate 10,000 variant orbits for each best-fit orbit to construct our \emph{Uncertainty Tube}.} These three case studies are also shown in greater detail in the supplemental material accompanying this paper. The domain expert selected these asteroids to represent a wide range of challenging scenarios that the visualizations of \neoviz~could address:

\noindent \textbf{Close approach~(\cref{sec:2012DA14})\;} A close approach with Earth drastically alters a NEO's orbital parameters, increasing the uncertainty in unpredictable ways as the resulting changes depend on the distance between the object and Earth.

\noindent \textbf{Predicted impactor~(\cref{sec:99942Apophis})\;} Predicting the location many decades into the future is challenging as its covariance matrix quickly becomes unbounded without further observations.  This use case illustrates \neoviz's ability to inspect the time-evolution of impactors in 3D, investigate the uncertainty in location over 25 years of orbits, and display an \emph{Impact Map} for the small likelihood of an impact.

\noindent \textbf{Imminent impactor~(\cref{sec:2023CX1})\;} Many NEOs are only discovered hours or days before their impact.  This use case demonstrates \neoviz's ability to visualize an object with such a small number of observations and to aid the inspection of historical uncertainty.

\subsection{Close Approach of (367943) Duende}
\label{sec:2012DA14}
\begin{figure}[t]
  \centering
  \includegraphics[width=\columnwidth]{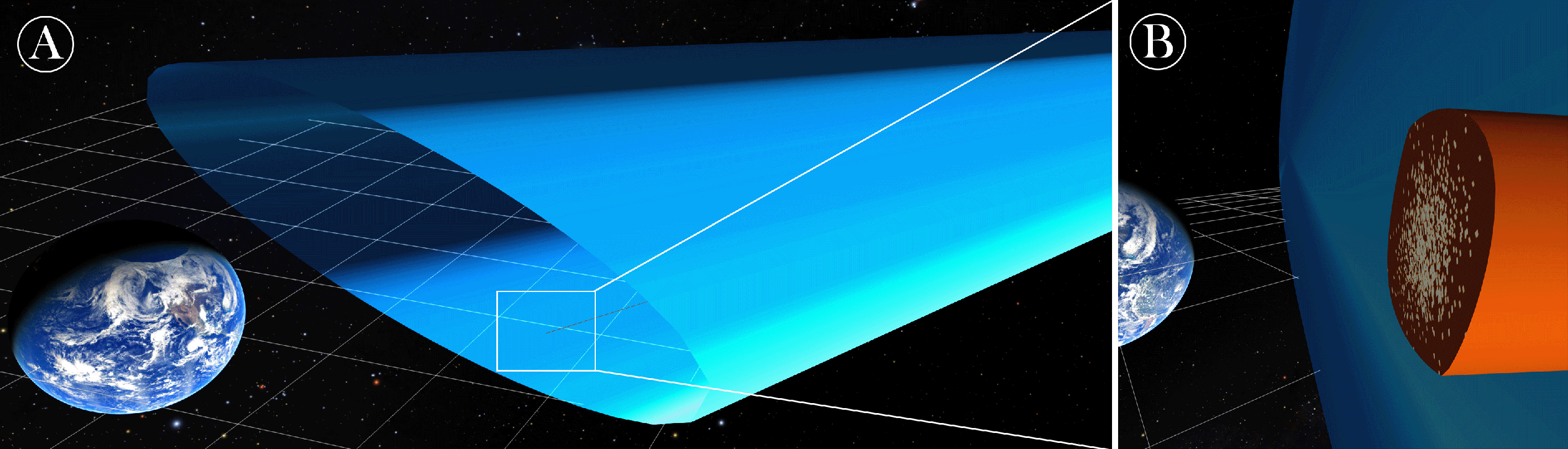}
  \caption{Nested \emph{Uncertainty Tube}s for asteroid (367943) Duende as propagated from submissions on March 5\dateth, 2012 (blue), and February 10\dateth, 2013 (orange), respectively. (B) is a zoom-in view of (A).  The distance between grid cells is 7,500\,km.}
  \label{fig:da14-two-tubes}
\end{figure}

Asteroid (367943) Duende is a small near-Earth asteroid that made a record-close approach on February 15\dateth, 2013.  The flyby was so close that it traversed inside the geosynchronous satellites, reaching the closest distance of about 27,700 kilometers from Earth's surface.  Using \neoviz~, we visualize the changes in the uncertainty of orbit prediction as the number of observations increases and show the progression of the close approach in relation to the associated uncertainty.

\cref{fig:da14-two-tubes} shows the recreation of the asteroid orbit predictions at two different times.  We use the \emph{historical uncertainty} representation of the tube and focus on two submissions, one on March 5\dateth, 2012, one month after its discovery, and one on February 10\dateth, 2013, five days before the close approach.  The orbit trajectories are propagated forward from each of these submissions to an end time after the close approach.  Two nested \emph{Uncertainty Tube}s are generated.  \cref{fig:da14-two-tubes} (A) shows both tubes shortly before the close approach.  The larger blue tube represents the trajectory uncertainty predicted using data from March 5\dateth, 2012.  The uncertainty region is much larger than Earth with a relatively high probability of intersection.  The smaller orange tube represents an uncertainty region, computed using more recent observations, that would approach but not intersect Earth.  We also observe a more circular cut-plane and more uniformly distributed points due to a decrease in uncertainty, indicating a more confident estimation of the asteroid trajectory.  \neoviz~effectively showcases the substantial reduction in uncertainty as the number of observations increases.

To show the progression of the close approach, we generate a \emph{sectioned uncertainty} tube for the asteroid using all known observations.
\edit{We compute a \emph{positional uncertainty magnitude} at a given time, using the covariance matrix.  We consider the first quadrant of the covariance matrix, which only represents the positional variance. The three dimensions correspond to the Cartesian coordinates of the asteroid $(x, y, z)$. The diagonal of the quadrant, $(\sigma_x, \sigma_y, \sigma_z)$ represents the variance in each of the dimensions, respectively. We compute the positional uncertainty magnitude as the geometric length of that diagonal.  This uncertainty magnitude approximates the radius of the uncertainty region which is correlated to the longest axis of the \emph{Uncertainty Tube}.}
\cref{fig:da14-close} (A) is a dual-axis plot displaying the inverse correlation between the uncertainty magnitude in blue and the number of observations in orange.  The three timestamps before, during, and after the close approach are marked in orange on the uncertainty magnitude, which is also shown in \cref{fig:da14-close} (B), (C), and (D), respectively, where the yellow trails show the geostationary satellites.  The uncertainty is large before the close approach due to a lack of observations.  We observe an interesting phenomenon on the cut-plane of the \emph{Uncertainty Tube}, where most trajectories accumulate on the side of the tube in the direction of the Earth, indicating the asteroid's movement approaching Earth.  The uncertainty decreases as the asteroid moves closer to Earth.  The \emph{Uncertainty Tube} reduces in size and the cut-planes circularize as the distance uncertainty decreases drastically with increased numbers of observations.  This perception corresponds exactly to the findings in the uncertainty plot.  The \emph{Uncertainty Tube} enables a visual inspection of the changes in asymmetric uncertainty in more detail.

\subsection{Historical High Probability Impact of (99942) Apophis} \label{sec:99942Apophis}
\begin{figure}[b!]
  \centering
  \includegraphics[width=\columnwidth]{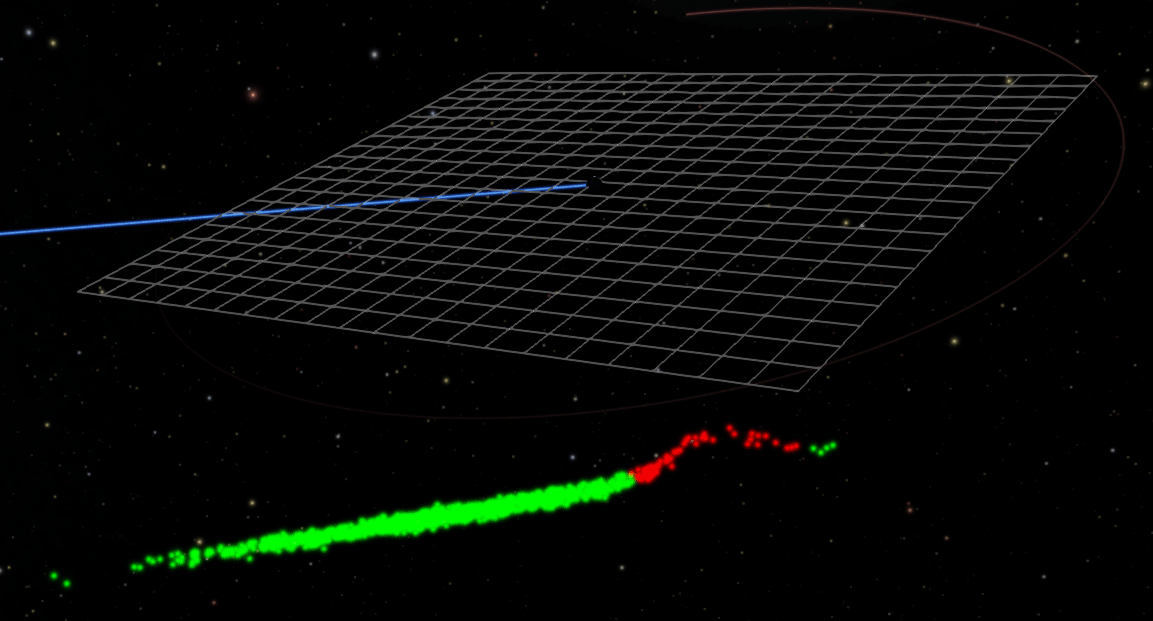}
  \caption{A point cloud visualization showing 1000 possible trajectories for Apophis, moments before impact on April 13\dateth, 2029. The impactor trajectories, marked in red, are visibly more affected by Earth's gravity.}
  \vspace*{-\baselineskip}
  \label{fig:point-cloud}
\end{figure}

For this case study, we recreate a historical prediction of the high probability impact event of the asteroid (99942) Apophis, estimated to be 300 to 400 meters along its longest axis.  Upon its discovery in June 2004, it quickly gained notoriety as it was recognized as potentially hazardous based on the observations of December 20\dateth, 2004, where JPL predicted the asteroid's potential impact on Friday, April 13\dateth, 2029 with an impact probability of 0.02\%.  However, the impact probability continued to increase, peaking on December 27\dateth~at a 2.7\% likelihood of impact in 2029~\cite{Chesley2005}.  Although the impact in 2029 was later ruled out following further observations, it remains a significant historical event with potentially daunting implications.  We recreate this high-impact probability prediction using \neoviz~to show the possible trajectories of Apophis given the available data on December 27\dateth, 2004 (see \cref{fig:point-cloud}).

\begin{figure}[b!]
  \centering
  \subfloat[Uncertainty magnitude]{\includegraphics[height=3.6cm]{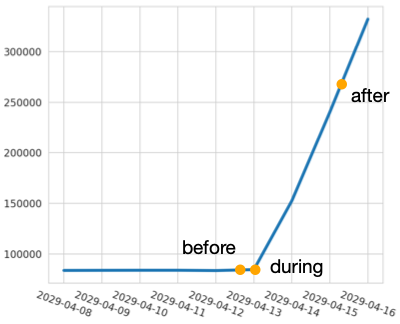}}
  \hfill
  \subfloat[Impactor vs non-impactor samples]{\includegraphics[height=3.6cm]{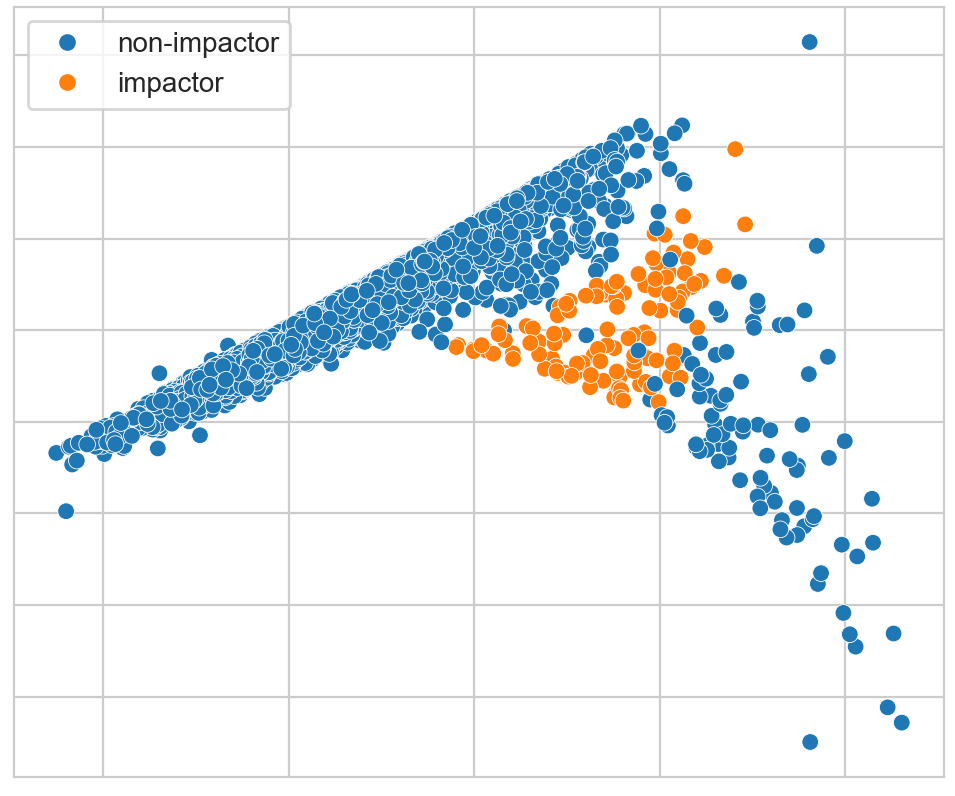}}
  \caption{The uncertainty magnitude for Apophis drastically increases after the predicted impact in 2029.  The cut-plane shows the cluster of samples that will impact (orange) and miss (blue) Earth.}
  \label{fig:mn4-cutplane}
\end{figure}

\begin{figure*}[t]
  \centering
  \subfloat[Before close approach]{\includegraphics[width=0.33\linewidth]{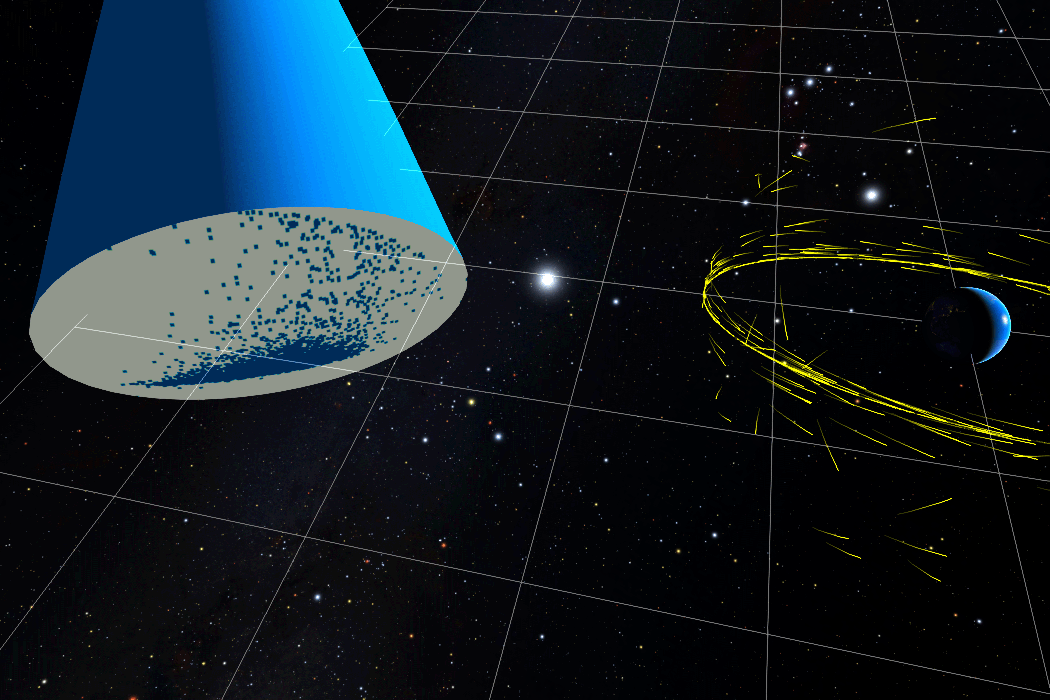}}
  \hfill
  \subfloat[During close approach]{\includegraphics[width=0.33\linewidth]{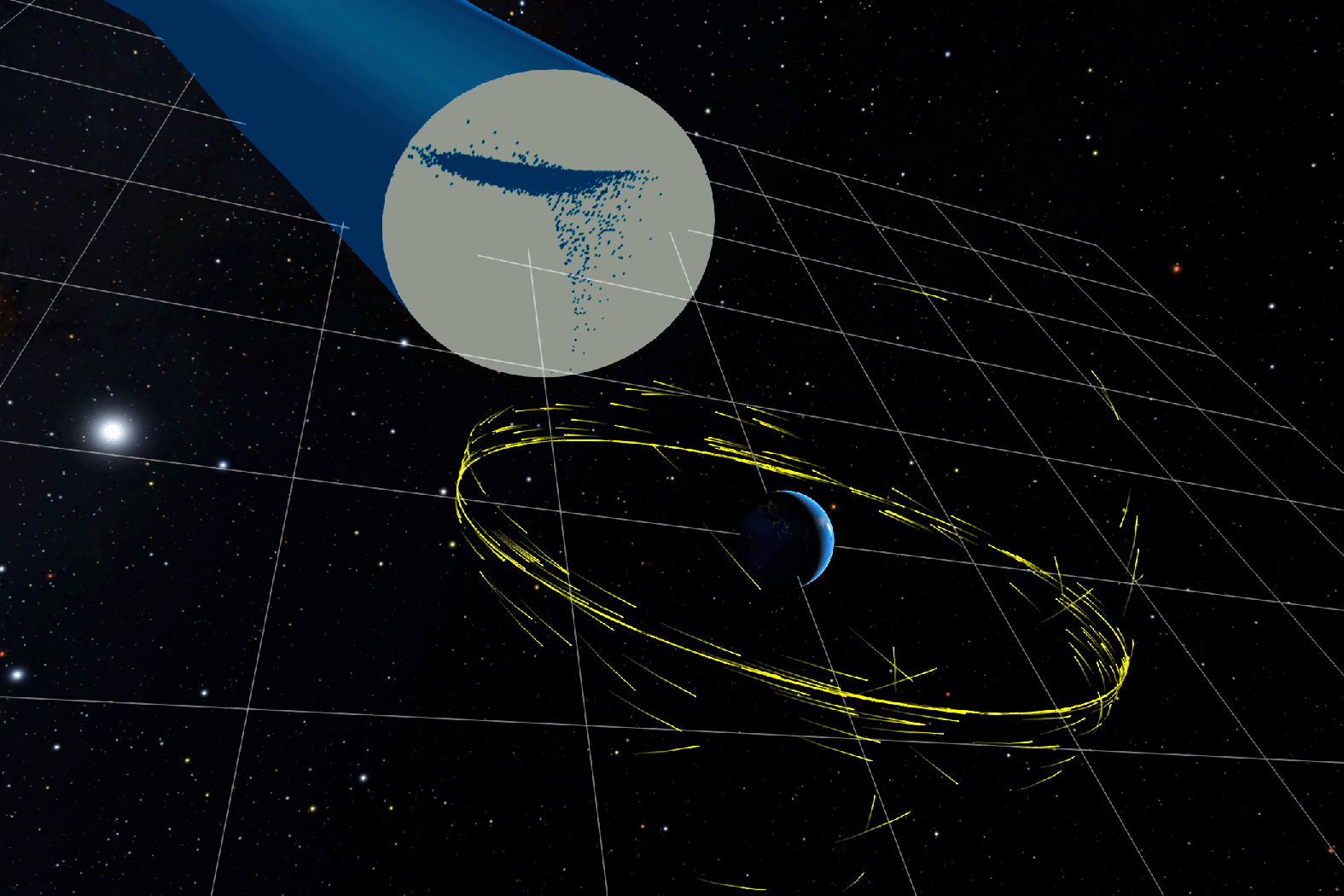}}
  \hfill
  \subfloat[After close approach]{\includegraphics[width=0.33\linewidth]{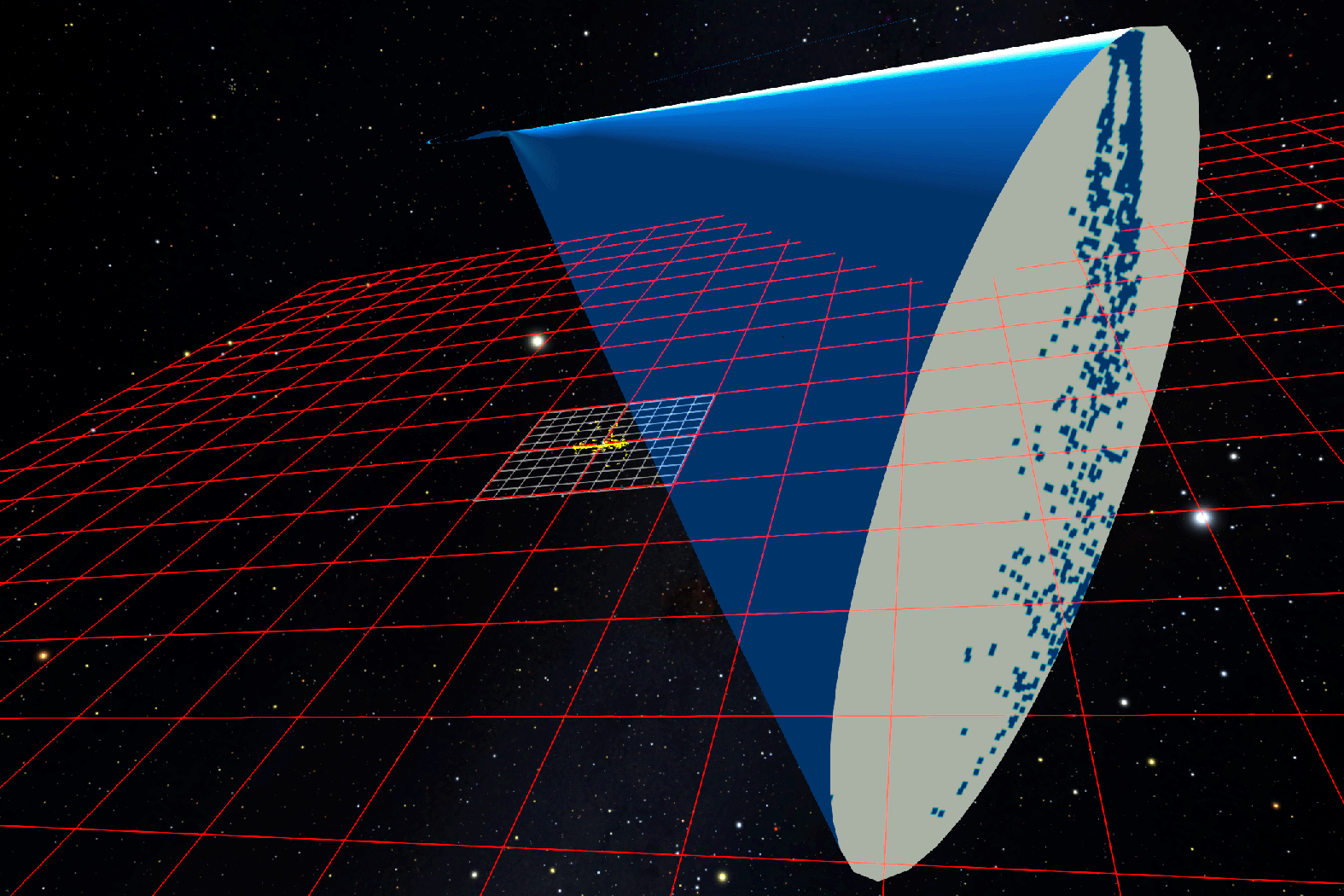}}
  \caption{(99942) Apophis impact event on April 13\dateth, 2029 at time 21:43:59 UTC.  Note that the data used to generate this image comes from observational data in 2004 when the impact probability for the object was at its highest.  This does not reflect what will happen in 2029 as additional observations since 2004 have ruled out an impact.  Due to the gravitational influence of Earth, the distribution of trajectory samples inside the \emph{Uncertainty Tube} shifts as it approaches Earth.  The close encounter causes a rapid increase of the uncertainty (see also~\cref{fig:mn4-cutplane}, A).  The grid lines of the gray reference grid have a size of 30,000\,km.  The yellow lines around Earth represent the geostationary satellites at a distance of 36,000\,km.}
  \vspace*{-\baselineskip}
  \label{fig:apophis-close}
\end{figure*}

\begin{figure}[b!]
  \centering
  \includegraphics[width=\linewidth]{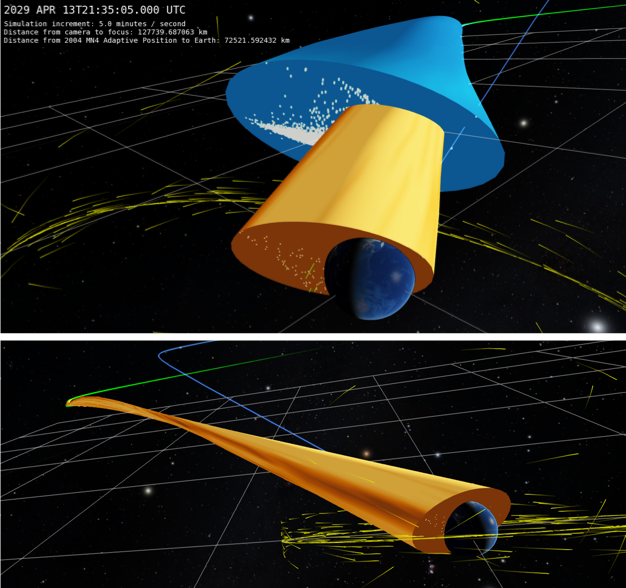}
  \caption{The top image combines two \emph{Uncertainty Tube}s to show a combination of all trajectories (blue) and only impactors (orange). The \emph{Uncertainty Tube} is wider than Earth as impacting trajectories can lead or lag behind the mean velocity plane and thus arrive at Earth at different times. The bottom image displays the impactor tube only.}
  \label{fig:mn4-intersection}
\end{figure}

We compute a \emph{historical uncertainty} tube for Apophis from the last submission on December 27\dateth, 2004 to shortly after the predicted impact time on April 13\dateth, 2029.  We sample 10,000 trails from the uncertainty distribution at the initial submission and propagate forward.  We observe a slight decrease in uncertainty hours before the predicted impact time and an immediate and drastic increase in uncertainty after approaching Earth.  Shown both in the uncertainty magnitude plot in~\cref{fig:mn4-cutplane} (A) and the \neoviz~visualization in~\cref{fig:apophis-close}. The \emph{Uncertainty Tube} decreased in diameter from (A) to (B) but increased in size significantly from (B) to (C).  Out of these 10,000 trails, we identify 115 trails that impact Earth, indicating an impact probability of 1.15\%, comparable to the historical record~\cite{Chesley2005}.  We generate a second \emph{Uncertainty Tube} using only the impact trails. We observe in~\cref{fig:mn4-intersection} (top) that at the same point in time, the impactor tube is further ahead while the larger tube lags behind. This indicates an interesting phenomenon that the impactor trajectories traveled further on average than the non-impactor trajectories.
As we move forward in time to the predicted impact time around 21:00 UTC on April 13\dateth, 2029, we show both a nested tube and the impactor tube alone, see~\cref{fig:mn4-intersection}. We witness a small elevation in the sub-tube of the impactor trajectories, representing the perturbation in the asteroid orbit caused by Earth's gravitational pull.

\begin{figure}[b!]
  \centering
  \vspace{-6mm}
  \includegraphics[width=0.75\columnwidth]{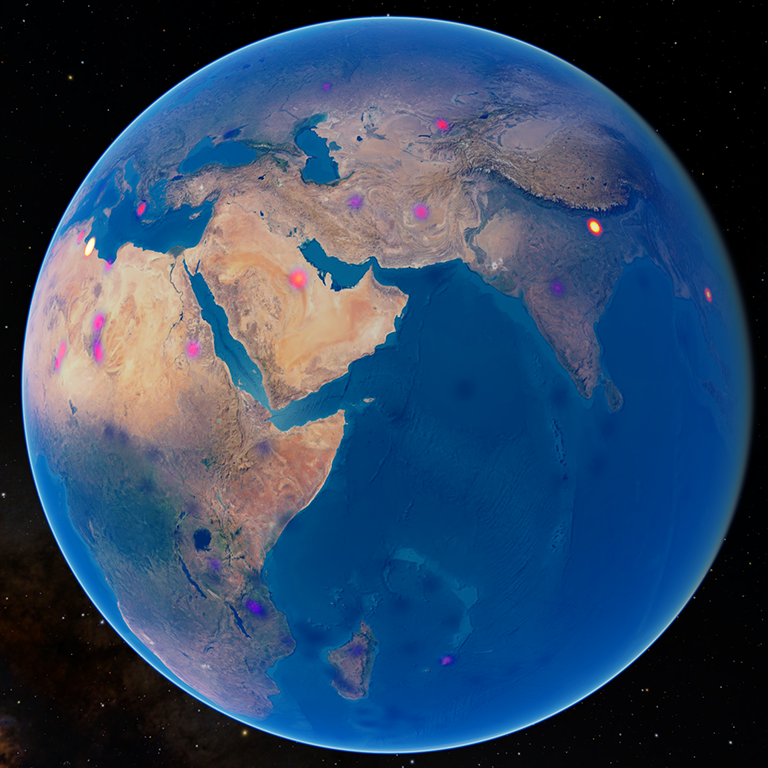}
  \caption{Risk assessment for (99942) Apophis in 2029 showing the convolution between the likelihood of an impact with population density.}
  \vspace*{-\baselineskip}
  \label{fig:mn4-map-impact}
\end{figure}

We further investigate the cause of this phenomenon. Since the point distribution on the cut-planes is reflected in the surface color of the \emph{Uncertainty Tube}, we can change the color map on the surface to reveal the changes in the point distribution more clearly over time, as shown in the large blue tube in \cref{fig:teaser}. We observe that the surface colors change drastically approximately 2 days before the impact date. At the time of the predicted impact, the points on the cut-plane diverged from an elliptical Gaussian to the plot shown in \cref{fig:mn4-cutplane} (B). We see that the impact trails (orange) are clustered together in relation to all sampled trails (blue). This change in shape of the impactor tube suggests that the asteroid is near Earth for some time before the impact, allowing the Earth's gravity to have a large effect on its orbit. We can also deduce that the asteroid has low impact velocity or $v_\infty$  before it was accelerated by Earth's gravity. Our finding is confirmed by JPL's Small-Body Database\footnote{\url{https://ssd.jpl.nasa.gov/tools/sbdb_lookup.html\#/}~}, where the $v_\infty$ for Apophis on the impact date is predicted to be 5.84 compared to an average value of about 20. We show this effect more directly using a time-varying point cloud visualization of 1,000 trajectories, see \cref{fig:point-cloud}. The red points are the impactor trajectories and the green are regular trajectories. We see that the impactor trajectories are pulled towards Earth and diverge from the rest of the ensembles.

For all 115 potential impact trajectories for Apophis, we compute their impact locations and generate an \emph{Impact Map} (see~\cref{sec:system}).  We present the impact probability (\cref{fig:teaser}) and the population risk views (\cref{fig:mn4-map-impact}).  In the impact probability view, we see many potential impact points in Africa, a few in the Indian Ocean, and Europe.  The brighter impact points, reflecting higher impact probability, are mostly located close to Madagascar and Tanzania.  In the population risk view, the brightness of the impact points is determined by the density of the population at the impact locations.  We see the brightest points in Tripoli, Patna, Riyadh, Tashkent, and Bangkok, corresponding to areas of high population.  This view offers compelling visual cues of urgency that could potentially aid in the development of evacuation strategies to mitigate disastrous outcomes of an asteroid impact.

\subsection{Imminent Impactor \designation{2023}{CX}{1}} \label{sec:2023CX1}
\begin{figure}[b!]
  \centering
  \vspace{-4mm}
  \includegraphics[width=\linewidth]{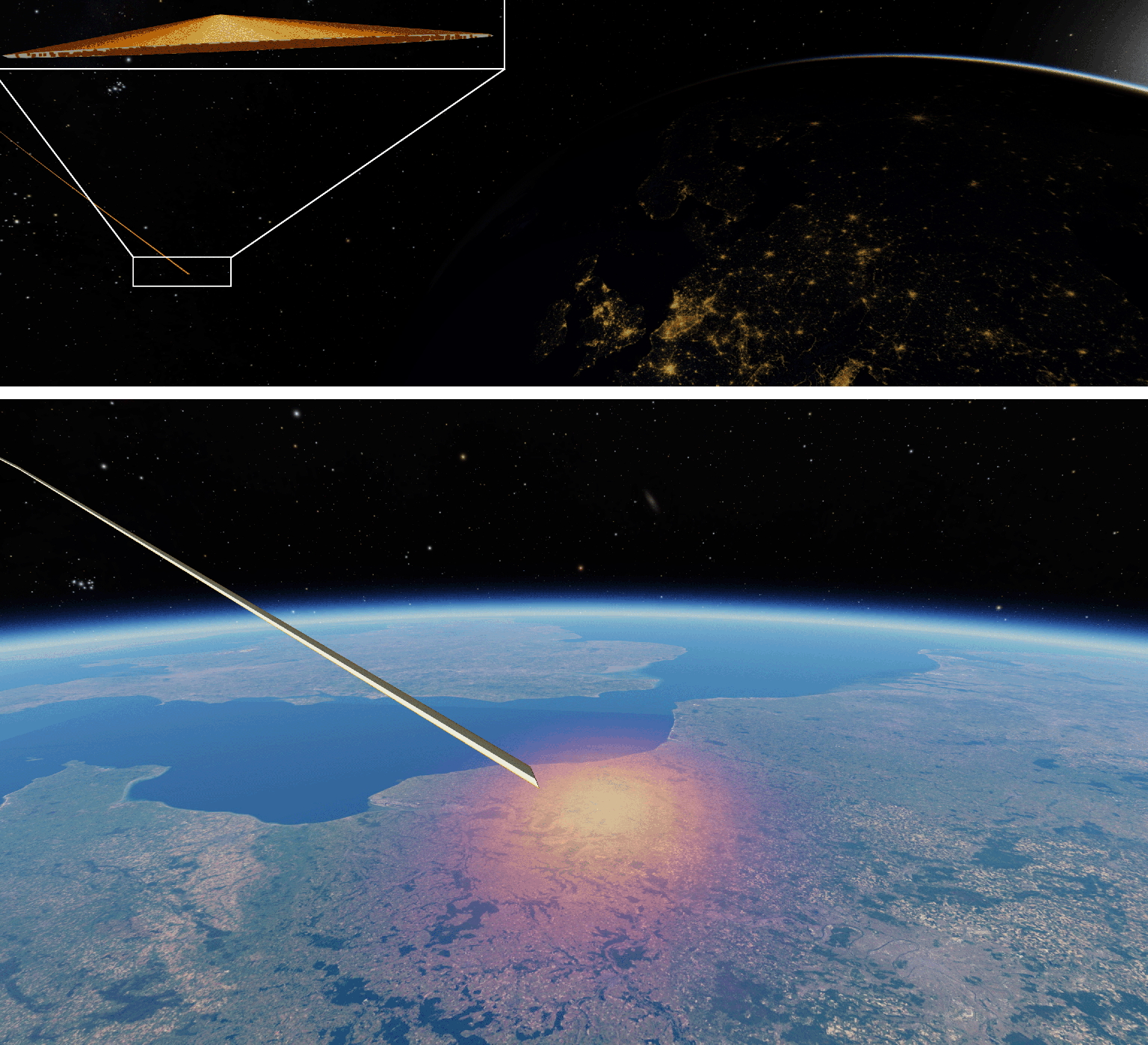}
  \vspace{-2mm}
  \caption{The imminent impactor \designation{2023}{CX}{1} on 2023 February 13 02:59:21 UTC is visualized with the \emph{Uncertainty Tube}.  The top image shows an orange tube that is highly asymmetrical.  The lower image displays the impact location.  As the impact occurred during the night, Earth was artificially lit to make the surrounding geography visible.}
  \vspace*{-\baselineskip}
  \label{fig:cx1-impact}
\end{figure}

\designation{2023}{CX}{1} was a small asteroid discovered at the Konkoly Observatory in Hungary on February 12\dateth, 2023 around 21:20 UTC that entered Earth's atmosphere over northwestern France six hours later at approximately 2:59 UTC.  The asteroid created a bright fireball across the English Channel, which was visible in France, the United Kingdom, Germany, and Belgium~\cite{bischoff2023saint}.  A previous visualization of the impact corridor provided by the European Space Agency displays a short band moving towards Normandy, France\footnote{\url{https://neo.ssa.esa.int/past-impactors/2023cx1}~}.  \neoviz~shows a comparable result, augmented with additional information and conveyed with enhanced visual impact.  Due to the extremely short window between initial discovery and impact, only a few hundred observations are available for \designation{2023}{CX}{1}. This results in a highly asymmetrical \emph{historical uncertainty} tube shown in~\cref{fig:cx1-impact}. Using all available knowledge, we show the tube from 02:38 to 03:40 on Feb 13\dateth, 2023.  As the asteroid is an imminent impactor, the impact location is constrained to a small region as the uncertainty does not have time to disperse over a larger surface area before impacting.  \cref{fig:cx1-impact} shows the \emph{Uncertainty Tube} and the \emph{Impact Map} created using \neoviz.  The \emph{Uncertainty Tube} has a narrow shape, representing a relatively high uncertainty in the depth direction.  Our predicted impact region is close to the coast in Normandy, France, aligning with the other predictions.  \neoviz~presents an interactive tool to simultaneously display the asteroid trajectory and the location of the imminent impactor.

%% file: 6-expert-feedback.tex
\section{Expert Feedback}\label{sec:expert-feedback}
The development of \neoviz~followed a participatory design process in collaboration with a domain expert from the B612 foundation. To more systematically assess the effectiveness of our visualization, we conducted a series of in-depth interviews with five experts, who had no prior experience with \neoviz. These experts were co-founder and director of the B612 Foundation, Dr. Ed Lu, and renowned researchers in asteroids and small solar system bodies, Dr. Aren Heinze, Dr. Mario Juri\'c, and Dr. Siegried Eggl. All experts have extensive expertise on NEOs and their associated uncertainty, as well as substantial experience using visualization in their research; one also has experience developing new visualization tools. During the interviews, we presented a tutorial of \neoviz, followed by an interactive demonstration of the Apophis case study. The experts then provided feedback on the effectiveness of each component of the visualization tool and offered suggestions for improvement. We refer to the experts as E1-E5, in no particular order.

We received positive feedback and encouragement for \neoviz, especially for its novel approach to visualizing uncertainty. E1 acknowledged that we are tackling a ``tough visualization task'', and remarked, \emph{``I haven't seen anything like this, and I think it is great''}. Both E3 and E4 discussed the difficulties they face in visualizing trajectory uncertainty, noting that \neoviz~is a unique tool with the potential to ``be a game changer'' for the planetary defense community.

For the \emph{Uncertainty Tube} and the cut-planes, E1 commented that they are superior to the point cloud visualization as one can clearly see the shape of the complex 3D structure in a time-varying setting. E3 viewed the cut-planes as an effective way to ``provide a 2D snapshot of a fairly complex 3D structure'' and appreciated the changing of tube shape as new observations arrived. E2 and E4 emphasized that the distribution of the points is useful for further analysis. E2 found the elevation in the tube before the Apophis potential impact ``fascinating'', saying that ``we need visualization like this to see'' such features. However, experts also provided valuable suggestions for improvement. E1 and E2 expressed reservations about removing the temporal aspects from the visualization, as the cut-planes depict the projection of the trajectories instead of their true locations. E2 proposed adding an option for a cut-surface that respects the true time of each trajectory, in addition to the cut-planes. E4 suggested providing a statistical view on the cut-planes by showing a kernel density estimation map of the point distribution. 

Experts responded positively towards the impact map, with E4 stating it was an excellent idea to convolute population density on top of the impact locations. E1 suggested offering users more flexibility in adjusting the size, color, and brightness of the impact markers. E2 and E4 expressed interest in using an upsampling approach to form more defined impact corridors instead of scattered impact points. However, this is currently beyond our computational and rendering capabilities.

\neoviz~generated several surprising new insights and inspired many promising research directions. 
E2 and E3 were both fascinated by the elevation in the impactor tube, and expressed curiosity about the extent of trajectory scrambling over the 15-year propagation period. E3 found the point cloud visualization particularly engaging, especially where the impactor orbits are visibly pulled towards Earth, stating \emph{``that’s something you don’t often think about, and you obviously see it here.''} E3 also highlighted potential applications of \neoviz~in visualizing spacecraft trajectories, and the possibility of generalizing it to star orbits, while cautioning about associated complexities. E4 discussed applying \neoviz~to space collision problems, such as space debris and collisions within the asteroid belt. On a broader note, E2 appreciated \neoviz~as a tool for unscheduled exploration of asteroid data, stating, \emph{``I spend a lot of time staring at data and just exploring it. Sometimes it feels like a waste, but I think it is actually really productive in the long run. And this seems like a tool one can use for that kind of exploration.''}

%% file: 7-conclusion.tex
\section{Conclusion and Future Work} \label{sec:conclusion}
In this paper, we present {\neoviz}, a system enabling planetary defense experts to analyze the time-varying uncertainty in the trajectories of NEOs. Our approach introduces two key visualizations: the \emph{Uncertainty Tube} and the \emph{Impact Map}. The \emph{Uncertainty Tube} provides a novel time-varying 3D uncertainty representation of the asteroid trajectories, allowing users to explore the temporal evolution of the orbital uncertainty in context with other objects in the Solar System, such as Earth and satellites. The \emph{Impact Map}, on the other hand, displays potential impact locations, combined with population risk, on a 3D model of Earth using the \emph{GlobeBrowsing} system~\cite{vis17-bladin-globe-browsing}. By integrating these visualizations into a common reference frame, {\neoviz}~enables planetary defense experts to analyze the uncertainty evolution of NEO trajectories to an extent that was unattainable with previous techniques. {\neoviz}~addresses unique challenges in working with asteroid data, which exhibits large variability in both time and space due to the diversity in asteroid scenarios. We demonstrate the capabilities of {\neoviz}~with three asteroids: (367943) Duende, (99942) Apophis, and \designation{2023}{CX}{1}, and further evaluate our system with expert feedback interviews.

While the expert interviews showed that {\neoviz} has significant promise as a tool for analyzing asteroid trajectories, it has some limitations. Projecting the 3D point cloud onto a 2D plane results in information loss and inaccuracies in the uncertainty representation. Additionally, the current system does not support displaying the numerical data about the uncertainty tube or the asteroid. Addressing these limitations could be valuable directions for future work. Moreover, the expert interviews highlighted the scientific benefits of providing statistical views of tbe uncertainty representation by using kernel density estimation. The orbit variant generation process could also be further improved.

In light of new telescopes providing an abundance of asteroid data, we would like to extend the system to visualize multiple asteroids simultaneously, and to enable comparative analysis of NEOs on a population level rather than individually. Furthermore, we see the potential of using {\neoviz}~for public outreach. The system could be used to create immersive experiences in planetariums, raise awareness of NEOs, and ultimately garner public support for funding future missions aimed at safeguarding Earth from catastrophic impacts.

\acknowledgments{
This work was supported by the Knut and Alice Wallenberg Foundation under grant KAW 2019.0024 and through the WISDOME project, NASA under grant NNX16AB93, the Swedish e-Science Research Centre, the DIRAC Institute in the Department of Astronomy at the University of Washington, and the Asteroid Institute, a program of B612.  The DIRAC Institute is supported through generous gifts from the Charles and Lisa Simonyi Fund for Arts and Sciences, and the Washington Research Foundation.  The Asteroid Institute is supported through generous gifts from William K. Bowes Jr. Foundation, Tito's CHEERS, McGregor Girand Charitable Endowment, Galinsky Family, D. Kaiser, P. Rawls Family Fund, J \& M Montrym, Y. Wong and Kimberly Algeri-Wong and three anonymous leadership donors plus donors from 46 countries worldwide.

We would also like to thank Drs. Ed Lu, Aren Heinze, Mario Juri\'c, and Siegried Eggl for taking the time and giving us invaluable feedback during the expert interviews. Finally, we would also like to thank the reviewers for this paper that have helped improve this manuscript.

This visualization system was implemented in the \emph{OpenSpace} system~\cite{vis19-bock-openspace-system} and its source code is available at \texttt{https://github.com/OpenSpace/OpenSpace}.
}